\documentclass[preprint,11pt,review,authoryear]{elsarticle}

\usepackage{amssymb}
\usepackage[table]{xcolor}
\usepackage{amsmath}
\usepackage{amsthm}
\usepackage{lineno}
\usepackage{comment}
\usepackage{float}
\usepackage{xcolor}
\usepackage{tabularx}
\usepackage{array}
\usepackage{multirow}
\usepackage{longtable}
\usepackage{makecell}
\usepackage{booktabs}
\usepackage{acronym}
\usepackage{chngcntr}
\usepackage{threeparttable} 
\usepackage{threeparttablex}
\usepackage{footnote} 
\usepackage{longtable}  
\usepackage{caption}
\usepackage{lipsum}

\newcolumntype{C}{>{\centering\arraybackslash}X}
\begin{document}
\begin{frontmatter}

\title{Coupled two-phase flow and surfactant/PFAS transport in porous media with angular pores: From pore-scale physics to Darcy-scale modeling}

\author{Sidian Chen}
\cortext[mycorrespondingauthor]{Corresponding author}
\ead{sidianc@stanford.edu}

\address{Department of Energy Science and Engineering, Stanford University, USA}

\author{Bo Guo}

\address{Hydrology and Atmospheric Sciences, The University of Arizona, USA}

\author{Tianyuan Zheng}

\address{College of Environmental Science and Engineering, Ocean University of China, China}

\begin{abstract}

Two-phase surfactant-laden flow and transport in porous media are central to many natural and engineering applications. Surfactants modify two-phase flow by altering interfacial tension and wettability, while two-phase flow controls surfactant transport pathways and adsorption sites. These coupled processes are commonly modeled by combining Darcy-type two-phase flow equations with advection–dispersion–adsorption transport equations, with capillary pressure–saturation relationships scaled using the Leverett $J$-function. However, the Leverett $J$-function simplifies the porous medium as bundles of cylindrical tubes and decouples interfacial tension and wettability, limiting representation of angular pore geometries and coupled interfacial tension and wettability effects. We present a modeling framework that incorporates pore angularity and interfacial tension--wettability coupling effect into Darcy-scale surfactant-laden flow and transport models. Within this framework, we derive two-phase flow properties for angular pores, upscale them across pore size distributions, and obtain explicit and closed-form expressions for the upscaled properties. These expressions are incorporated into a coupled flow--transport model for simulating transient two-phase flow and surfactant transport processes. Results suggest a nonmonotonic and nonlinear dependence of two-phase flow properties on pore structure (angularity and size distribution) and interfacial tension (controlled by surfactant type and concentration). Example simulations of water flow and PFAS (surfactant-like contaminants) migration in unsaturated soils indicate that surfactant-induced flow effects on PFAS leaching are generally minor under typical site conditions, whereas pore angularity exerts dominant control on water flow, interfacial area, and consequently PFAS retention. Overall, the upscaling framework offers a more physically grounded approach for modeling two-phase surfactant-laden fluid flow and surfactant transport in porous media.

\end{abstract}


\end{frontmatter}


\section{Introduction}
\label{sec:intro}

Two-phase surfactant-laden fluid flow and surfactant transport in porous media play an important role in many natural and industrial applications. Surfactants can reduce interfacial tension between fluids and alter solid surface wettability through interfacial adsorption, which may alter fluid flow through porous structures. Fluid flow, in turn, controls the advection and diffusion of surfactants and governs the fluid-fluid and fluid–solid interfaces that facilitate the interfacial adsorption of surfactants. These mechanisms have been harnessed to remove pollutants (e.g., non-aqueous phase liquid (NAPL) and per- and polyfluoroalkyl substances (PFAS)) from soils and groundwater \citep[e.g.,][]{al2009impact,maroli2024surfactant}, enhance oil and gas recovery \citep[e.g.,][]{pope1978chemical,lake1989enhanced}, improve CO$_2$ and hydrogen storage efficiency \citep[e.g.,][]{foyen2020increased,chaturvedi2022air}, and optimize fluid behavior in manufactured porous materials such as cooling systems and CO$_2$ capture electrolyzers \citep[e.g.,][]{ge2022dynamically}. Consequently, mechanistic understandings and accurate predictions of the coupled two-phase flow and surfactant transport are essential for advancing technologies in these environmental, energy, and climate applications.

Modeling surfactant-laden fluid flow in porous media is challenging due to the nonlinear coupling between fluid flow and surfactant transport. A common strategy is to couple Darcy-scale two-phase flow with surfactant transport \citep[e.g.,][]{pope1978chemical,abriola1993surfactant,smith1994effect}, where flow is described by two-phase extended Darcy's law (or simplified forms such as Richards' equation) and transport is governed by advection--dispersion--adsorption equations. Most models did not account for surfactant adsorption at fluid-fluid interfaces. Surfactant effects are typically incorporated by scaling the capillary pressure--saturation relationship with the Leverett $J$-function (i.e., $J(S) \propto 1/(\gamma\cos\theta)$, where $S$ is fluid saturation, $\theta$ the contact angle, and $\gamma$ the fluid--fluid interfacial tension), while other flow properties such as relative permeability and interfacial area--saturation relations are usually assumed unchanged \citep[e.g.,][]{smith1994effect}.

The Leverett $J$-function, originally derived semi-empirically by \citet{leverett1941capillary}, relies on two major simplifications. First, it represents pore spaces as bundles of cylindrical tubes, whereas many porous media exhibit angular geometries that may strongly alter fluid configurations, capillary pressure, relative permeability, and interfacial area. Second, it assumes that interfacial tension and contact angle are independent, which may break down when surfactants are present. Surfactant adsorption modifies fluid--fluid and solid--fluid interfacial tensions, thereby altering the contact angle. Accounting for pore angularity and the coupling between interfacial tension and contact angle is critical for more accurate predictions of two-phase flow properties of different surfactant-free fluid systems \citep[e.g.,][]{tokunaga2013capillary,wang2016capillary}.

The impact of pore angularity on two-phase surfactant-free fluid flow is well established through experiments \citep[e.g.,][]{dong1995imbibition,oren2003reconstruction} and pore-scale and upscaling simulations \citep[e.g.,][]{tuller1999adsorption,diamantopoulos2013physically,diamantopoulos2015closed,jiang2020pore,jiang2020characterization,chen2021effect}. In contrast, two-phase flow of surfactant-laden fluids in angular porous media remains underexplored \citep{wijnhorst2020surfactant}. Existing studies rely mainly on experiments and empirical models \citep[e.g.,][]{desai1991influence,karagunduz2001influence}, which are difficult to generalize across pore structures, wettability conditions, surfactant chemistries, and concentrations. This highlights the need to develop predictive pore-scale and upscaling models for deriving two-phase flow properties under diverse conditions and improving coupled two-phase flow--surfactant transport frameworks.

Among the modeling approaches, the bundle-of-capillary-tubes model provides an attractive framework. It can efficiently quantify flow and transport properties at the representative elementary volume (REV) scale \citep[e.g.,][]{diamantopoulos2013physically,chen2021effect}, yielding explicit and/or closed-form expressions that can be directly integrated into Darcy-scale models \citep{diamantopoulos2015closed}. Moreover, it captures the effects of pore angularity for surfactant-free fluids with good agreements with experimental observations \citep[e.g.,][]{or1999liquid,tuller2001hydraulic,diamantopoulos2013physically,jiang2020characterization}. Building on this foundation, we extend the bundle-of-capillary-tubes concept to develop an upscaling workflow for deriving REV-scale two-phase flow properties of surfactant-laden fluids in angular porous media and integrating the derived properties into Darcy-scale coupled flow and transport models. This upscaling workflow consists of four key steps: (1) compute surfactant-laden fluid configurations in a single angular pore represented by a angular tube; (2) derive the upscaled two-phase flow properties for a porous medium whose void spaces are represented by a bundle of tubes; (3) formulate explicit and closed-form expressions for the new properties; and (4) integrate these expressions into Darcy-scale flow and surfactant transport models. This is the first framework that explicitly integrates pore angularity and interfacial tension--contact angle coupling into Darcy-scale two-phase flow surfactant transport models. 

\begin{figure}[!htb]
  \centering
  \includegraphics[width=1\textwidth]{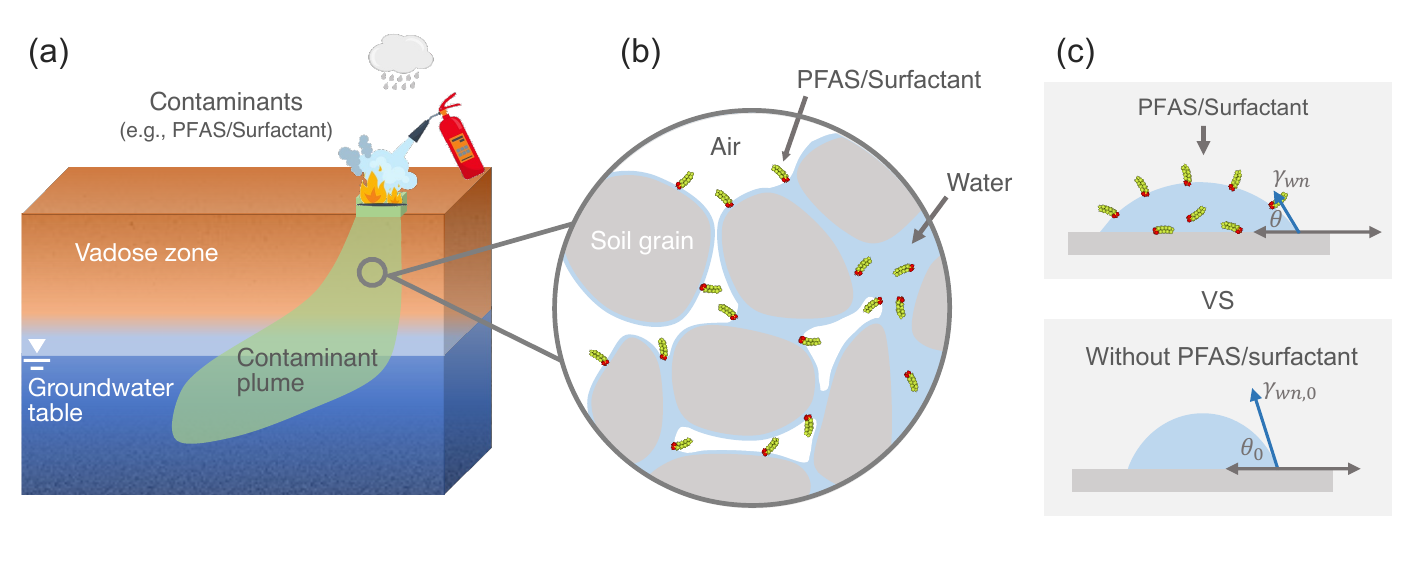}
  \caption{(a) Schematic of soil contamination by surfactant-like chemicals, such as per- and polyfluoroalkyl substances (PFAS). (b) Accumulation and adsorption of PFAS at the air--water interfaces in unsaturated soil pores. (c) Surface tension and contact angle of PFAS-laden water compared with PFAS-free water. Panels (a) and (b) are revised from \citet{chen2023pore} with permission of the authors and Wiley.}
  \label{fig:concept}
\end{figure}

We demonstrate the workflow's potential by applying it to a representative case: PFAS transport in unsaturated soils. PFAS, a class of fluorinated surfactant contaminants, have raised global health concerns due to their ubiquity, persistence, bioaccumulation, and toxicity at ng/L concentrations. Predicting their migration in soils remains challenging because it is governed by the coupled effects of interfacial interactions, surface tension variations, and transient unsaturated flow within irregular, often angular, pore spaces (Figure~\ref{fig:concept}). Existing numerical models---typically based on Richards’ equation coupled with advection--dispersion--adsorption formulations---account for surface tension effects only through scaling the capillary pressure--saturation curve using the standard Leverett $J$-function \citep[e.g.,][]{guo2020mathematical,silva2020modified,zeng2021multidimensional}. These approaches inherently neglect the influence of pore angularity and the coupling between interfacial tension and contact angle, both of which may strongly affect two-phase flow properties and PFAS transport. We employ our framework to quantify how these factors regulate PFAS migration in unsaturated soil columns under laboratory conditions, providing insights for future field-scale modeling efforts.

\section{Methods}
\label{sec:model}

\subsection{Interfacial tension and contact angle for surfactant-laden fluids} \label{subsec:IFT-theta}

For two fluids (i.e., a wetting-phase fluid and a nonwetting-phase fluid) that are surfactant-free and resting on a solid surface, the interfacial forces at the three-phase contact line follow Young–Dupré equation \citep{young1805essay,dupre1869theorie}
\begin{equation}
   \gamma_{wn,0} \cos \theta_{0} = \gamma_{sn,0} - \gamma_{sw,0},
\end{equation}
where $\gamma_{wn,0}$, $\gamma_{sn,0}$, and $\gamma_{sw,0}$ are the interfacial tension between surfactant-free fluids, the interfacial tension between solid and surfactant-free nonwetting-phase fluid, and the interfacial tension between solid and surfactant-free wetting-phase fluid, respectively; $\theta_0$ is the intrinsic contact angle between the solid and surfactant-free fluids. 

When surfactants are present, they can adsorb at the fluid--fluid and fluid--solid interfaces, altering their interfacial tensions. Because the interfacial forces at the three-phase contact line remain balanced, we obtain
\begin{equation}
   \gamma_{wn} \cos \theta = \gamma_{sn} - \gamma_{sw},
   \label{eq:sigma-theta}
\end{equation}
where $\gamma_{wn}$, $\gamma_{sn}$, and $\gamma_{sw}$ are the interfacial tension between surfactant-laden solid and surfactant-laden nonwetting-phase fluid, the interfacial tension between surfactant-laden solid and surfactant-laden wetting-phase fluid, and the interfacial tension between surfactant-laden fluids, respectively; and $\theta$ is the contact angle in the presence of interfacial adsorption of surfactants. Note that if $\gamma_{wn}$ > $|\gamma_{sn} - \gamma_{sw}|$, the surface is partially wet to the wetting-phase fluid. Once $\gamma_{wn}$ becomes smaller than $|\gamma_{sn} - \gamma_{sw}|$, the contact line disappears and the solid surface is completely wet by the wetting-phase fluid.

In Equation (\ref{eq:sigma-theta}), $\gamma_{wn}$ can be given by the Szyszkowski equation,
\begin{equation}
   \gamma_{wn} = \gamma_{wn,0}\left[1 - b_\alpha \ln\left(1 +\frac{C_\alpha}{a_\alpha}\right)\right],
   \label{eq:Szyszkowski}
\end{equation}
where $C_\alpha$ is the surfactant concentration in the wetting-phase fluid ($\alpha = w$) or nonwetting-phase fluid ($\alpha = nw$), and $a_\alpha$ and $b_\alpha$ are model parameters obtained via fitting to measured surface tension data. Accordingly, we can compute the fluid--fluid interfacial excess ($\Gamma_{wn}$) via the Gibbs equation (i.e., $\Gamma_{wn} = - \frac{1}{R_gT} \frac{\partial \gamma_{wn}}{\partial \ln C_{\alpha}}$, where $R_g$ is the universal gas constant, $T$ is temperature), which yields $\Gamma_{wn}  = \gamma_{wn,0} b_\alpha C_{\alpha} /[R_gT(a_\alpha + C_{\alpha})]$. Additionally, assuming the solid--fluid interfacial adsorption follows the Freundlich isotherm (i.e., $\Gamma_{s\alpha} = K_{f,\alpha} C_{\alpha}^{N_{f,\alpha}}$, where $\gamma_{s\alpha}$ is the surfactant excess at the fluid--solid interfaces, $K_{f,\alpha}$ and $N_{f,\alpha}$ are fitting parameters with experimental measurements), we can derive a Freundlich analog of the Szyszkowski equation for $\gamma_{SN}$ and $\gamma_{SW}$ by plugging the Freundlich isotherm and Gibbs equation and integrating with respect to the surfactant concentration,
\begin{equation}
   \gamma_{s\alpha} = \gamma_{s\alpha,0} - \int_{0}^{C_{\alpha}} R_g\, T\, \Gamma_{s\alpha}(C)\, {\rm d} \ln C = \gamma_{s\alpha,0} - \frac{R_g T  K_{f,\alpha}} {N_{f,\alpha}} C_{\alpha}^{N_{f,\alpha}},
   \label{eq:Freundlich}
\end{equation}
where $\Gamma_{s\alpha}$ is the surfactant excess at the interface between solid and wetting-phase fluid ($\alpha = w$) or at the interface between solid and nonwetting-phase fluid ($\alpha = nw$). It is worth noting that Equations~(\ref{eq:Szyszkowski}) and (\ref{eq:Freundlich}) predict monotonic decreases in $\gamma_{wn}$ and $\gamma_{s\alpha}$ as surfactant concentration increases. In practice, these interfacial tensions often level off above a characteristic concentration (e.g., the critical micelle concentration). Capturing this behavior requires modified expressions (e.g., asymptotic forms), which are not considered in the present study.

\begin{figure}[!htb]
  \centering
  \includegraphics[width=1\textwidth]{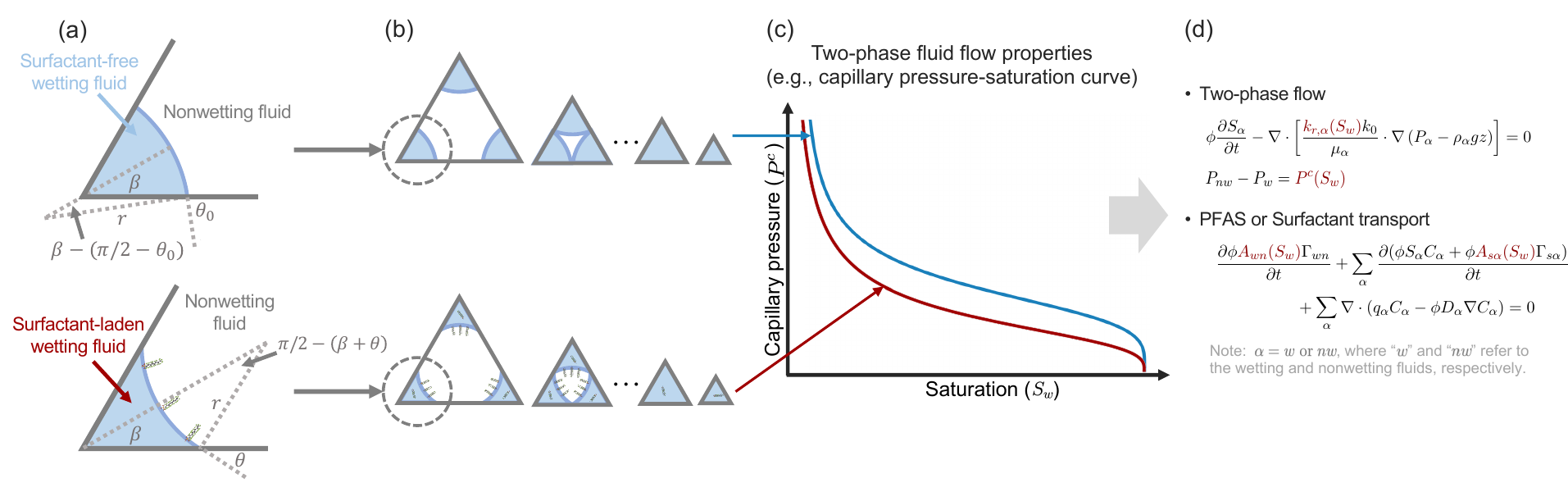}
  \caption{A upscaling workflow to bridge pore-scale physics of coupled two-phase flow and surfactant/PFAS transport into a Darcy-scale modeling framework. The workflow consists of four steps: (a) Compute the configuration of the wetting and nonwetting fluids in an angular corner for a given capillary pressure with or without the presence of surfactant/PFAS in the fluids. Equilateral triangular tubes are used as examples. (b) Compute the configuration of the wetting and nonwetting fluids in a bundle of capillary tubes for a given capillary pressure. (c) Apply the bundle-of-capillary-tubes model to derive two-phase flow properties (e.g., capillary pressure, relative permeability, and fluid--fluid interfacial area vs. fluid saturation curves) for a porous medium. The capillary pressure vs. wetting-phase fluid saturation curve is used as an example. (d) Couple the new two-phase flow properties into Darcy-scale transient two-phase flow and surfactant/PFAS transport models. Note that a nanometer-scale thin wetting-phase fluid film (referred to as ``precursor film'') may form on partially-wet surfaces, while a macroscopic thin wetting-phase fluid film (thicker than precursor film) will form on completely-wet surfaces. Due to the small film thickness, the thin films are not shown in Panels (a) and (b).}
  \label{fig:workflow}
\end{figure}

Because $\gamma_{wn}$ and $\theta$---which collectively control the capillary pressure and fluid configuration in the pore spaces---are functions of surfactant concentrations, the two-phase flow properties (i.e., the relationships among capillary pressure, relative permeability, and fluid--fluid interfacial area, and fluid saturation) will depend on the surfactant concentrations. We derive the two-phase flow properties for surfactant-laden fluids using a bundle-of-capillary-tubes model (Figure \ref{fig:workflow}a--c). In particular, the model represents the complex pore structures in a porous medium by a bundle of capillary tubes with idealized geometries. Following this simplified representation, we can derive two-phase flow properties for surfactant-laden fluids in an individual pore of a porous medium and upscale the individual-pore relationships for the porous medium with an arbitrary pore size distribution. The specific procedures are explained in the following sections

\subsection{Two-phase flow properties in a pore with varying geometries} \label{subsec:cons-props-a-pore}

We first derive the two-phase flow properties for surfactant-laden fluids in an individual pore with idealized geometries (e.g., cylindrical, square, and triangular tubes).

\subsubsection{Capillary pressure} \label{subsubsec:pc-s-a-pore}
We start from a pore that is saturated by the wetting-phase fluid and is connected to a reservoir of a nonwetting-phase fluid. As we increase the reservoir pressure, the nonwetting-phase fluid will invade the pore when the pressure difference between nonwetting- and wetting-phase fluids (i.e., the capillary pressure $p^c$) exceeds a threshold (referred to as ``critical capillary pressure $p^c_{cr}$'' hereafter). In a cylindrical pore,
$p^c_{cr}$ is given by \citep{dullien2012porous}
\begin{equation}
    p^c_{cr} = \frac{\gamma_{wn} \cos \theta} {2 R},
    \label{eq:pc-cyl}
\end{equation}
where $R$ is the radius of the circular cross-section of the cylindrical pore. In an angular pore, $p^c_{cr}$ is assumed equal to the $p^c$ where the fluid-fluid interfaces meet in the angular pore, which is given by
\begin{equation}
   p^c_{cr} = \frac{\gamma_{wn}}{r_c} 
\label{eq:pc-rc}
\end{equation}
where $r_c$ is the radius of curvature when the menisci meet and collapse at the edges of the cross-sections of a pore. The expression of $r_c$ depends on the pore geometries. In a square-tube pore, $r_c$ is given by \citep{chen2020fully}
\begin{equation}
    r_c =  \frac{ R\sin(\pi/4)}{|\sin(\pi/4-\theta)|},
    \label{eq:pc-squ}
\end{equation}
where $R$ is the inscribed radius of the cross-section of the square-tube pore. The $r_c$ in a triangular-tube pore is given by
\begin{equation}
r_c = \min_{\substack {i, j \in \{1,2,3\} \\ \&\; i \neq j}}\left( \frac{ R (\cot \beta_i + \cot \beta_j) }{|\sin(\pi/2-\beta_i-\theta)|/ \sin \beta_i + |\sin(\pi/2-\beta_j -\theta)|/ \sin \beta_j } \right),
\label{eq:pc-tri}
\end{equation}
where $\beta_i$ and $\beta_j$ refer to $i^{th}$ and $j^{th}$ half corner angles, respectively; $R$ is the inscribed radius of the cross-section of the triangular-tube pore.

When $p^c \le p^c_{cr}$ in a pore, the pore is saturated by the wetting-phase fluid. Once $p^c$ exceeds $p^c_{cr}$, the nonwetting-phase fluid invades the pore. The wetting-phase fluid will either reside at the corners of angular pores (referred to as ``corner'' fluid hereafter) or form thin fluid films on the solid surfaces not covered by bulk or corner fluids. Two types of thin fluid films may form. In a partially-wet pore, the wetting-phase fluid may form a nanometer-scale thin film on the surface (referred to as ``precursor'' film hereafter) due to surface adsorption, capillary condensation, surface roughness, or surface heterogeneity. In a completely wet pore, the wetting-phase fluid will spread on the surface and form a macroscopic thin film  (referred to as ``macroscopic'' film hereafter) that is thicker than the precursor film. Because the films are often extremely thin, their volumes are almost negligible (if they exist) compared with the total fluid volume, we neglect their contribution to the overall saturation. Accordingly, the wetting-phase fluid saturation in an invaded cylindrical pore is assumed to be zero, while that in an invaded angular pore is given by
\begin{equation}
s_w = \frac{r^2  \tilde{A}_c }{ R^2 \sum_{i=1}^{N_{\beta}}  \cot\beta_i},
\label{eq:pc2s-pore}
\end{equation}
where $r$ is the radius of meniscus curvature given by $r = \gamma_{wn}/p^c$, $N_{\beta}$ is the number of corners of the cross-section of the angular pore (e.g., $N_{\beta} = 3$ in a triangular-tube pore and $N_{\beta} = 4$ in a square-tube pore), $\beta_i$ is the half-corner angle, and $\tilde{A}_c$ is the dimensionless cross-sectional area of a pore with $r = 1$ and $\tilde{A}_c$ is given by
\begin{equation}
    \tilde{A}_c = \sum_{i=1}^{N_{\beta}} \left[ \frac{\sin(\pi/2-\beta_i-\theta)}{\sin \beta_i} \cos\theta - (\pi/2-\beta_i-\theta) \right].
\end{equation}
The nonwetting-phase fluid saturation is given by 
\begin{equation}
s_{nw} = 1 - s_w.
\end{equation}

\subsubsection{Relative permeability} \label{subsubsec:kr-s-a-pore}

If a pore is saturated by a fluid, the permeability is given by \citep{patzek2001verification, patzek2001shape}
\begin{equation}
k_{\alpha} = \eta G A_{\alpha}^2,
\label{eq:ksat}
\end{equation}
where $\eta$ is the corrector varying with pore geometry ($\eta = 0.5, 0.5623,$ or $ 0.6$ for cylindrical, square-tube, and equal-lateral triangular-tube pores, respectively), $G$ is the shape factor of the pore cross-section (i.e., the area divided by perimeter square), $A_{\alpha}$ is the area of fluid $\alpha$ in the cross-section of a pore, and $\alpha$ indicates the fluid of interest (i.e., $\alpha = w$ indicates the wetting-phase fluid, while $ \alpha = nw$ indicates the nonwetting-phase fluid). 

If two fluids coexist in a pore, the wetting-phase fluid will reside either as thin films in cylindrical pores or as corner fluid and thin films in angular pores. We assume the thin film permeability is negligible. In a cylindrical pore invaded by the nonwetting-phase fluid, the wetting-phase fluid permeability is 0, while the nonwetting-phase fluid permeability is given by Equation (\ref{eq:ksat}). In an angular pore invaded by nonwetting-phase fluid, the corner fluid contributes a non-negligible permeability, which is given by \citep{patzek2001verification}
\begin{equation}
k_{w} = 2\sum_{i=1}^{N_{\beta}} k_{w, i} = 2\sum_{i=1}^{N_{\beta}} \tilde{g}_{w,i} l^4_{w,i},
\label{eq:kw}
\end{equation}
where $k_{w, i}$ is the wetting-phase fluid permeability in a half corner with an angle of $\beta_i$, $l_{w,i}$ is the wetting-phase meniscus-apex distance along the wall at the $i^{th}$ corner (i.e., $l_{w,i} = r|\sin(\pi/2 - \beta_i - \theta)|/\sin \beta_i$), and $\tilde{g}_{w}$ is the dimensionless conductance of the wetting phase at the half corner ($\beta_i$) with a unit meniscus-apex distance (i.e., $l_{w,i}=1$). The nonwetting-phase fluid residing in the middle of the pore is given by \citep{qin2019dynamic}
\begin{equation}
k_{nw} = \eta G A_{nw}^2 s_{nw}^2.
\label{eq:knw}
\end{equation}

\subsubsection{Fluid--fluid interfacial area}\label{subsubsec:aff-s-a-pore}

In a cylindrical pore, fluid--fluid interfaces will arise from thin films. Assuming the surfaces are smooth, the fluid--fluid interfacial area $a_{wn}$ in an invaded cylindrical pore with a unit length is given by
\begin{equation}
a_{wn} = 2 \pi R,
\end{equation}
where $R$ is the radius of the pore that is invaded by nonwetting-phase fluid. In an angular pore, fluid--fluid interfaces will arise from both corner water and thin films. Because the thin-film fluid--fluid interfaces are comparable or even larger than those associated with the corner water, they should be accounted for in the calculation of $a_{wn}$. Assuming the surfaces are smooth, the fluid--fluid interfacial area $a_{wn}$ in an angular pore  with a unit length is given by
\begin{equation}
a_{wn} = 2 \sum_{i=1}^{N_{\beta}} \left( r \cdot |\pi/2 - \beta_i - \theta| + R - l_{w,i}\right),
\label{eq:awn}
\end{equation}
where $l_{w,i}$ is the wetting-phase meniscus-apex distance along the wall at the $i^{th}$ corner and can be calculated via Equation (\ref{eq:kw}).

\subsection{Two-phase flow properties in a porous medium} \label{subsec:cons-props-medium}

We then upscale the individual-pore two-phase flow properties for a porous medium through a bundle-of-capillary-tubes model. Assuming the pores in a porous medium share the same intrinsic contact angle ($\gamma_{wn,0}$) and the pore sizes follow a distribution given by $f(R)$ with $R$ being the inradius of the cross-section of a pore and $f$ being the probability density of pores whose cross-sectional inradius is equal to $R$, the upscaled two-phase flow properties in a porous medium can be computed via an integration over the pore size distribution. 

In particular, we compute the wetting-phase fluid saturation ($S$) as a function of capillary pressure ($P^c$) for a porous medium by integrating the individual-pore saturation over the pore size distribution, which yields
\begin{equation}
S = \frac{ \int f(R) \cdot v(R) \cdot s (R, P^c, \gamma_{wn}) \: {\rm d}R } { \int f(R) \cdot v(R) \: {\rm d}R },
\label{eq:pc-sw}
\end{equation}
where $v$ is the pore volume and $s$ is computed given that $p^c = P^c$.

Similarly, we compute the relative permeability under a macroscopic capillary pressure $P^c$ (and correspondingly saturation $S$) by taking the ratio between the cumulative permeability for a fluid in each pore and the permeability for the entire porous medium under saturated conditions, which is given by
\begin{equation}
k_{r,\alpha} = \frac{ \int f(R) \cdot  k_{\alpha}(R, P^c, \gamma_{wn}) \: {\rm d}R } { \int f(R) \cdot  k_0(R) \: {\rm d}R },
\label{eq:kr-sw}
\end{equation}
where $s$ is the saturation of the fluid of interest, which is a function of $R$, $P^c$, and $\gamma_{wn}$ for a given $\gamma_{wn,0}$ and $\theta_0$. Finally, we can combine Equations (\ref{eq:pc-sw}) and (\ref{eq:kr-sw}) to compute the $k_{r}$-$S$ curves.

Similarly, we compute the fluid--fluid interfacial area per unit pore volume under a macroscopic capillary pressure $P^c$ (and correspondingly saturation $S$) by integrating those in all pores, which yields 
\begin{equation}
A_{wn} =\frac{ \int f(R) \cdot a_{wn}(R, P^c, \gamma_{wn})  \: {\rm d}R } { \int f(R) \cdot A_0 (R) \: {\rm d}R },
\label{eq:AFF-sw}
\end{equation}
We can also combine Equations (\ref{eq:pc-sw}) and (\ref{eq:AFF-sw}) to compute the $A_{wn}$-$S$ curves.

While Equations (\ref{eq:pc-sw}--\ref{eq:AFF-sw}) provide the upscaled two-phase flow properties in porous media with arbitrary pore size distributions, simplified mathematical representations are desired to couple them into two-phase flow and transport models. The following section introduces their simplified mathematical representations under idealized conditions. 

 \subsection{Simplified mathematical representations of two-phase flow properties}\label{subsec:expression-function}

We consider a porous medium with uniform intrinsic contact angles, uniform pore geometries (i.e., cylinders, square tubes, or triangular tubes with angle-angle similarity), and log-normal pore size distributions. Accordingly, we can derive explicit and closed-form expressions for the two-phase flow properties.

\subsubsection{Explicit expressions} \label{subsubsec:scaling}

We drive the explicit functions for capillary pressure, relative permeability, and fluid-fluid interfacial area in the porous medium.

\paragraph{Capillary pressure-saturation curve}
We first derive the scaling function for the capillary pressure in a porous medium with a log-normal pore size distribution. Suppose $R_{\rm max}$ is the maximum size of the saturated pores that are not invaded. Then the saturation is given by
\begin{equation}
\begin{split}
    S = \frac{\tilde{A}_0 \int_0^{R_{\rm max}} f(R) R^2 \: {\rm d}R +  \tilde{A}_c r^2 \int_{R_{\rm max}}^{\infty} f(R) \: {\rm d}R }{ \tilde{A}_0\int_0^{\infty} f(R) R^2 \: {\rm d}R }
    \label{eq:pc-sw-anal}
\end{split}
\end{equation}
where $\tilde{A}_0$ is the dimensionless cross-sectional area of a pore with $R = 1$ (i.e., $\tilde{A}_0 = \pi$ for a cylindrical pore, $\tilde{A}_0 = 4$ for a square-tube pore, $\tilde{A}_0 = \sum_{i=1}^{3} \cot{\beta_i}$ for a triangular-tube pore), $A_c$ is the area of the fluid in the corner of an unsaturated angular pore and $A_c = 0$ if the pore is represented by a cylindrical tube, and $f(R)$ is the probability density function (PDF) of the pore sizes. $f(R)$ is given by
\begin{equation}
f(R) = \frac{1}{R \sigma \sqrt{2\pi}} \exp\left(-\frac{(\ln R - \mu)^2}{2\sigma^2}\right),
\end{equation}
where $\mu$ and $\sigma$ are the mean and standard deviation of the natural log of the pore size $R$.

To evaluate $S$, we need to compute $\int_0^{R_{\rm max}} f(R) R^2 \: {\rm d}R$ and $\int_{R_{\rm max}}^{\infty} f(R) \: {\rm d}R$ from the following equation,
\begin{equation}
\begin{split}
\int_0^{R_{\rm max}}  f(R) R^n\, dR &
=  e^{n\mu + \frac{1}{2}n^2 \sigma^2} \cdot \Phi\left(\frac{\ln R_{\rm max} - (\mu + n\sigma^2)}{\sigma}\right), \\
 \int_{R_{\rm max}}^{\infty} f(R)  R^n \: {\rm d}R & =  e^{n\mu + \frac{1}{2}n^2 \sigma^2} \cdot \left[ 1 - \Phi\left(\frac{\ln R_{\rm max} - (\mu + n\sigma^2)}{\sigma}\right)\right],
\end{split}
\label{eq:intergralRn}
\end{equation}
where $n$ denotes the exponent (power) of $R$ within the integrand, and $\Phi$ is the cumulative distribution function of the standard normal distribution.

\paragraph{Relative permeability-saturation curve}
The wetting-phase relative permeability is given by
\begin{equation}
    k_{r, w} = \frac{ \eta G \tilde{A}_0^2 \int_0^{R_{\rm max}} f(R)  R^4 \: {\rm d}R + \left( 2\sum_{i=1}^{N_{\beta}} \tilde{g}_{1,i} l^4_{1,i}\right) \int_{R_{\rm max}}^{\infty} f(R) \: {\rm d}R }{ \eta G \tilde{A}_0^2 \int_0^{\infty} f(R) R^4 \: {\rm d}R },
    \label{eq:kr1-sw-anal}
\end{equation}
while that of the non-wetting phase is given by
\begin{equation}
    k_{r, nw}  = \frac{  \tilde{A}_0^2  \int_{R_ {\rm max}}^{\infty} f(R)  R^4  \: {\rm d}R  -   2\tilde{A}_0 \tilde{A}_c r^2 \int_{R_{\rm max}}^{\infty} f(R)  R^2 \: {\rm d}R  +  \tilde{A}_c^2 r^4\int_{R_{\rm max}}^{\infty} f(R) \: {\rm d}R }{  \tilde{A}_0^2 \int_0^{\infty} f(R) R^4 \: {\rm d}R },
    \label{eq:kr2-sw-anal}
\end{equation}
where $\int_0^{R_{\rm max}} f(R)  R^4 \: {\rm d}R$, $\int_{R_{\rm max}}^{\infty} f(R) \: {\rm d}R$, $\int_{R_{\rm max}}^{\infty} f(R)  R^2 \: {\rm d}R$, and $\int_{R_{\rm max}}^{\infty} f(R)  R^4 \: {\rm d}R$ are given by Equation (\ref{eq:intergralRn})

\paragraph{Fluid-fluid interfacial area-saturation curve} 
Finally, we can compute the specific fluid-fluid interfacial area (i.e., the fluid-fluid interfacial area normalized by the total pore volume) for cylindrical tubes by
\begin{equation}
    A_{wn}  = \frac{ 2 \pi \int_{R_ {\rm max}}^{\infty} f(R)  R  \: {\rm d}R }{  \tilde{A}_0 \int_0^{\infty} f(R) R^2 \: {\rm d}R},
    \label{eq:A12-cylinder-sw-anal}
\end{equation}
and that for angular tubes by
\begin{equation}
    A_{wn}  = \frac{ \left(a^c_{wn} - 2\sum_{i=1}^{N_{\beta}} l_{w,i}\right) \int_{R_{\rm max}}^{\infty} f(R) \: {\rm d}R + 2 N_{\beta}\int_{R_ {\rm max}}^{\infty} f(R)  R  \: {\rm d}R }{  \tilde{A}_0 \int_0^{\infty} f(R) R^2 \: {\rm d}R},
    \label{eq:A12-angular-sw-anal}
\end{equation}
where $\int_{R_{\rm max}}^{\infty} f(R) \: {\rm d}R$ and $\int_{R_{\rm max}}^{\infty} f(R) R \: {\rm d}R$ are given by Equation (\ref{eq:intergralRn}).

Equations (\ref{eq:pc-sw-anal}) and (\ref{eq:kr1-sw-anal}--\ref{eq:A12-angular-sw-anal}) provide the explicit expressions for the two-phase flow properties in porous media with a uniform wettability, uniform pore geometry, and lognormal pore size distribution.

\subsubsection{closed-form expressions} \label{subsubsec:closed-form-funs} 

 The explicit expressions are given as functions of the cumulative normal distribution functions. We can approximate the cumulative normal distribution functions by closed-form expressions such as a logistic function $\Phi(x) \approx 1/[1 + \exp(-1.702x)]$. This approximation allows for obtaining closed-form expressions for the upscaled two-phase flow properties and conveniently coupling them into transient two-phase flow models.

\subsection{Coupled Darcy-scale two-phase flow and surfactant/PFAS transport model} \label{subsec:DarcyModel}

We introduce the transient two-phase flow and surfactant transport models that couple the new two-phase flow properties, as well as the numerical algorithm for solving the model.

\subsubsection{Two-phase flow model} \label{subsubsec:flow}

We describe the two-phase flow by an extended two-phase Darcy model,
\begin{equation}
    \phi \frac{\partial S_{\alpha}}{\partial t} + \nabla \cdot q_{\alpha}  = 0
    \label{eq:2p-flow}
\end{equation}
where $q_{\alpha} = - k_{r,\alpha}k_0 \mu_{\alpha}^{-1} \cdot\nabla \left(P_{\alpha} - \rho_\alpha g z\right)$, $k_{r,\alpha}$ are computed via Equations (\ref{eq:kr1-sw-anal}--\ref{eq:kr2-sw-anal}), $k_0$ is the absolute permeability, $P_{\alpha}$ is the pressure of $\alpha$ phase, $P_{\alpha}$ is constrained by $P_n - P_w = P^c$ with $P^c$ being given by Equation (\ref{eq:pc-sw-anal}), $z$ is the spatial coordinate (assuming positive downward).

\subsubsection{Surfactant and PFAS transport model} \label{subsubsec:transport}

The surfactants can dissolve in the fluids, migrate with the fluid flow via advection and dispersion, and adsorb onto or desorb from the fluid--fluid and fluid--solid interfaces. We consider a single-component surfactant system and instantaneous interfacial adsorption. Subsequently, the governing equation for surfactant transport is given by 
\begin{equation}
     \frac{\partial \phi A_{wn}\Gamma_{wn}}{\partial t} +  \sum_{\alpha} \left[\frac{\partial ( \phi S_{\alpha} C_{\alpha}  +  \phi A_{s\alpha}\Gamma_{s\alpha})}{\partial t} +\nabla\cdot \left(q_{\alpha}C_{\alpha}  - \phi S_{\alpha} D_{\alpha}\nabla C_{\alpha}\right)\right] = 0,
    \label{eq:surf-transport}
\end{equation}
where $D_{\alpha}$ is the dispersion coefficient in the $\alpha$ (i.e., $D_{\alpha} = \tau D_{0,\alpha} + \mathcal{L}_{\alpha} |q_{\alpha}/\phi /S_{\alpha}|$, where $\tau$ is the tortuosity which can be approximated as $\tau = (\phi S_{\alpha})^{7/3}/\phi^2$ \citep{millington1961permeability}, $D_{0,\alpha}$ is the molecular diffusion coefficient in free $\alpha$ phase, $\mathcal{L}_{\alpha}$ is the longitudinal dispersivity, $\Gamma_{wn}$ and $\Gamma_{s\alpha}$ are the surfactant excess at fluid--fluid and solid--fluid interfaces, respectively. $\Gamma_{wn}$ and $\Gamma_{s\alpha}$ are given by the Langmuir and Freundlich isotherms as discussed in Section \ref{subsec:IFT-theta}.

\subsubsection{Numerical algorithm} \label{subsubsec:algm}

Equations (\ref{eq:2p-flow}) and (\ref{eq:surf-transport}) represent the two-phase flow and surfactant transport models. The model has three primary unknown variables: $P_{\alpha}$, $P^c$ (or $S_{\alpha}$), and $C_{\alpha}$, where $\alpha = w$ or $nw$. They can be solved with well-imposed initial and boundary conditions. We solve the unknown variables using a backward-Euler finite difference method. At each time step, the resulting nonlinear discretized equations are solved iteratively using the Newton-Raphson method. The convergence of the iterations is accepted when the L$_\infty$-norm of the residuals and the L$_\infty$-norm of the updates of the primary variables (e.g., $P_{\alpha}$, $P^c$, and $C_{\alpha}$) are smaller than certain thresholds.

\section{Results and analysis}
\label{sec:results}

We apply the bundle-of-capillary-tubes model in Section \ref{sec:model} to analyze the impact of fluid-fluid interfacial tension and pore geometry on two-phase flow properties in porous media (Section \ref{subsec:Impact2Properties}). Then we use the model as a reference to evaluate the accuracy of the explicit and closed-form expressions of the upscaled two-phase flow properties (Section \ref{subsec:ExplicitEXPvsClosedForm}). Finally, we demonstrate an example application of a two-phase fluid flow and surfactant transport model that incorporates the newly-derived two-phase flow properties (Section \ref{subsec:model-couple}).

\subsection{Impact of interfacial tension and pore geometry on two-phase flow properties} \label{subsec:Impact2Properties}

\subsubsection{Numerical experiment design}
We model the impact of interfacial tension and pore geometry on two-phase flow properties using a porous medium whose pore sizes follow a log-normal distribution, with a mean pore size of $\mu = 100\,\rm{\mu m}$ and a normalized standard deviation of $\sigma = 0.3$ (Figure \ref{fig:psd}a). For illustration, we consider a near neutral-wet surface, characterized by an intrinsic contact angle of $\theta_0 = 80^{\circ}$. We assume that the solid-phase adsorption of surfactants and the alteration of solid surface chemistry are weak and negligible (i.e., $\gamma_{sn} = \gamma_{sn,0}$ and $\gamma_{sw} = \gamma_{sw,0}$, and consequently $\gamma_{wn}\theta = \gamma_{wn,0}\theta_0$). This assumption is adopted to focus on the effects of fluid–fluid interfacial tension and pore geometry, rather than reflecting a model limitation, since the current model accounts for solid-phase adsorption and its impact on wettability (Section~\ref{subsec:IFT-theta}). We simulate six $\gamma_{wn}$ values for cylindrical pores and nine $\gamma_{wn}$ values for angular pores, ensuring that the selected values correspond to the entire range of contact angles, i.e., $0 \leq \theta \leq \theta_{0}$ (Figures \ref{fig:cyl-pore}–\ref{fig:tri-pore}). Note that $\gamma_{wn}$ is also a proxy of the surfactant effect, i.e., a smaller $\gamma_{wn}$ corresponds to a more interfacially active surfactant and/or a higher surfactant concentration.

\subsubsection{Porous media with cylindrical pores}
\begin{figure}[!htb]
\vspace{-3pt}
  \centering
  \includegraphics[width=1\textwidth]{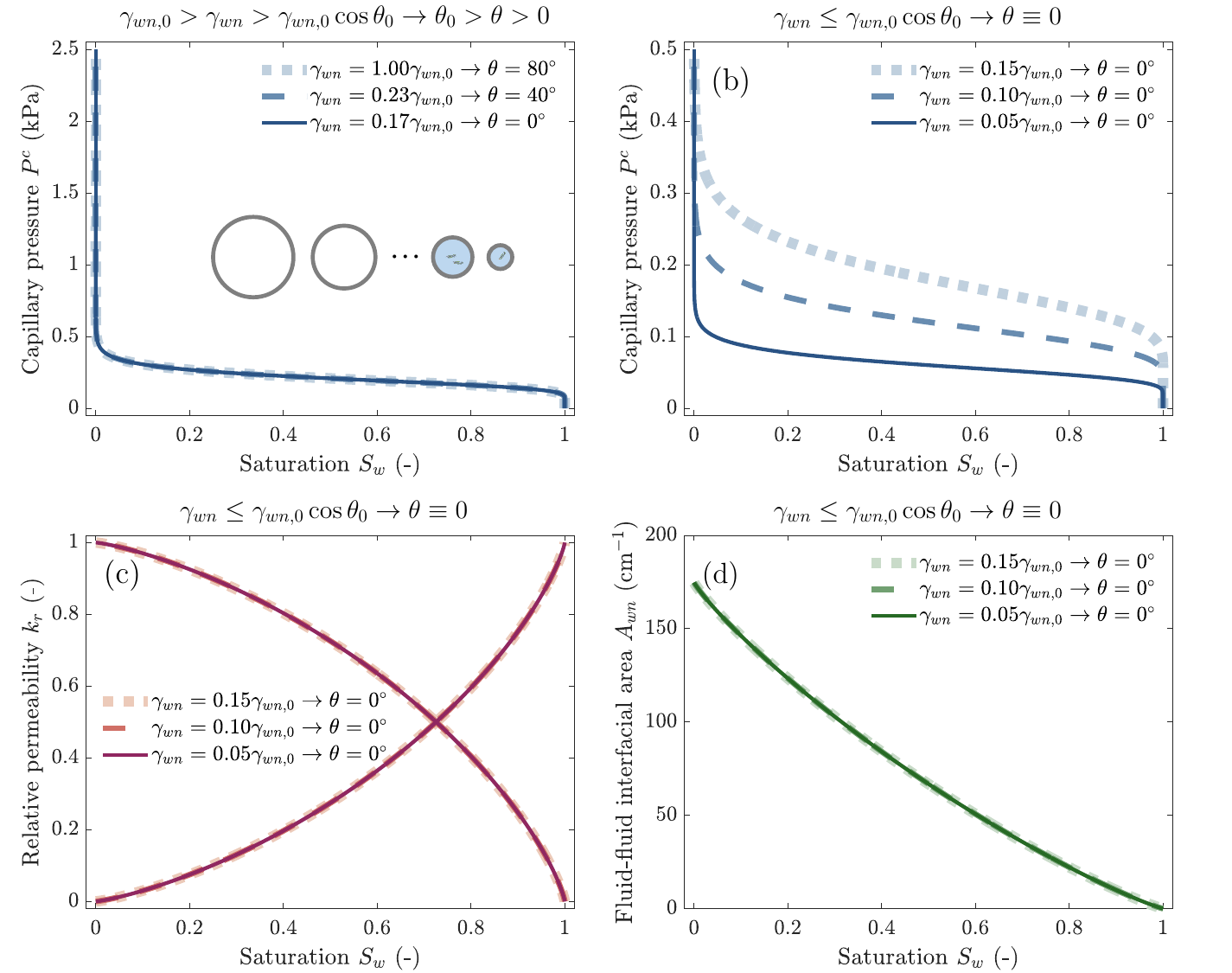}
  \caption{Impact of interfacial tension ($\gamma_{wn}$) on the (a--b) capillary pressure--saturation curves, (c) relative permeability--saturation curves, (d) fluid--fluid interfacial area--saturation curves in a porous medium with cylindrical pores. A near neutral-wet condition (e.g., an intrinsic contact angle $\theta_0$ of $80^{\circ}$) is used as an illustrative example. Two $\gamma_{wn}$ ranges are modeled: (1) A range within which the contact angles ($\theta$) are greater than or near 0, and (2) A range within which $\theta \equiv 0$. Because the relative permeability--saturation curves are the same in all cases, we only show those for contact angles greater than 0. In the figure, $\gamma_{wn,0}$ refers to the interfacial tension for surfactant-free fluids. Note that $\gamma_{wn}$ is a proxy of the surfactant effect, i.e., a smaller $\gamma_{wn}$ corresponds to a more interfacially active surfactant and/or a higher surfactant concentration.}
  \label{fig:cyl-pore}
\end{figure}

We first analyze the impact of $\gamma_{wn}$ on the two-phase flow properties in the porous medium with cylindrical pores. The results suggest a piecewise linear scaling of the capillary pressure--saturation ($P^c$--$S_w$) curves with $\gamma_{wn}$ (Figure \ref{fig:cyl-pore}a--b), which falls in the following two distinct regimes:

\begin{itemize}
    \item Regime I: $\gamma_{wn} > \gamma_{wn,0} \cos \theta_0$ (i.e., $\theta > 0$), corresponding to less interfacially active surfactants and/or lower surfactant concentrations. Because $\gamma_{wn} \cos \theta$ is constant in this regime and $P^c$ scales linearly with $\gamma_{wn} \cos \theta$ (Equation \ref{eq:pc-cyl}), the $P^c$--$S_w$ curves remain the same even $\gamma_{wn}$ varies (Figure \ref{fig:cyl-pore}a).

    \item Regime II: $\gamma_{wn} \le \gamma_{wn,0} \cos \theta_0$ (i.e., $\theta \equiv 0$), corresponding to more interfacially active surfactants and/or higher surfactant concentrations. Because $\cos \theta \equiv 1$ in this regime, the $P^c$--$S_w$ curves become linearly scaled with $\gamma_{wn}$ (Figure \ref{fig:cyl-pore}b).
\end{itemize}
In contrast, the relative permeability--saturation ($k_r$--$S_w$) and fluid--fluid interfacial area--saturation ($A_{wn}$--$S_w$) curves are the same for all $\gamma_{wn}$ (Figure \ref{fig:cyl-pore}c--d and Figure \ref{fig:pc-sw-cylinderical-pores}). In a bundle of cylindrical tubes, each tube will be occupied by one bulk fluid and a thin precursor film of the other fluid due to the absence of corner fluids. To reach a certain $S_w$, the partitioning of fluids across tubes is unique. Therefore, $k_r$ and $A_{wn}$ are kept the same at the same $S_w$ regardless of the varying $\gamma_{wn}$.

\subsubsection{Porous media with angular pores}
\begin{figure}[!htb]
  \centering
  \includegraphics[width=1.0\textwidth]{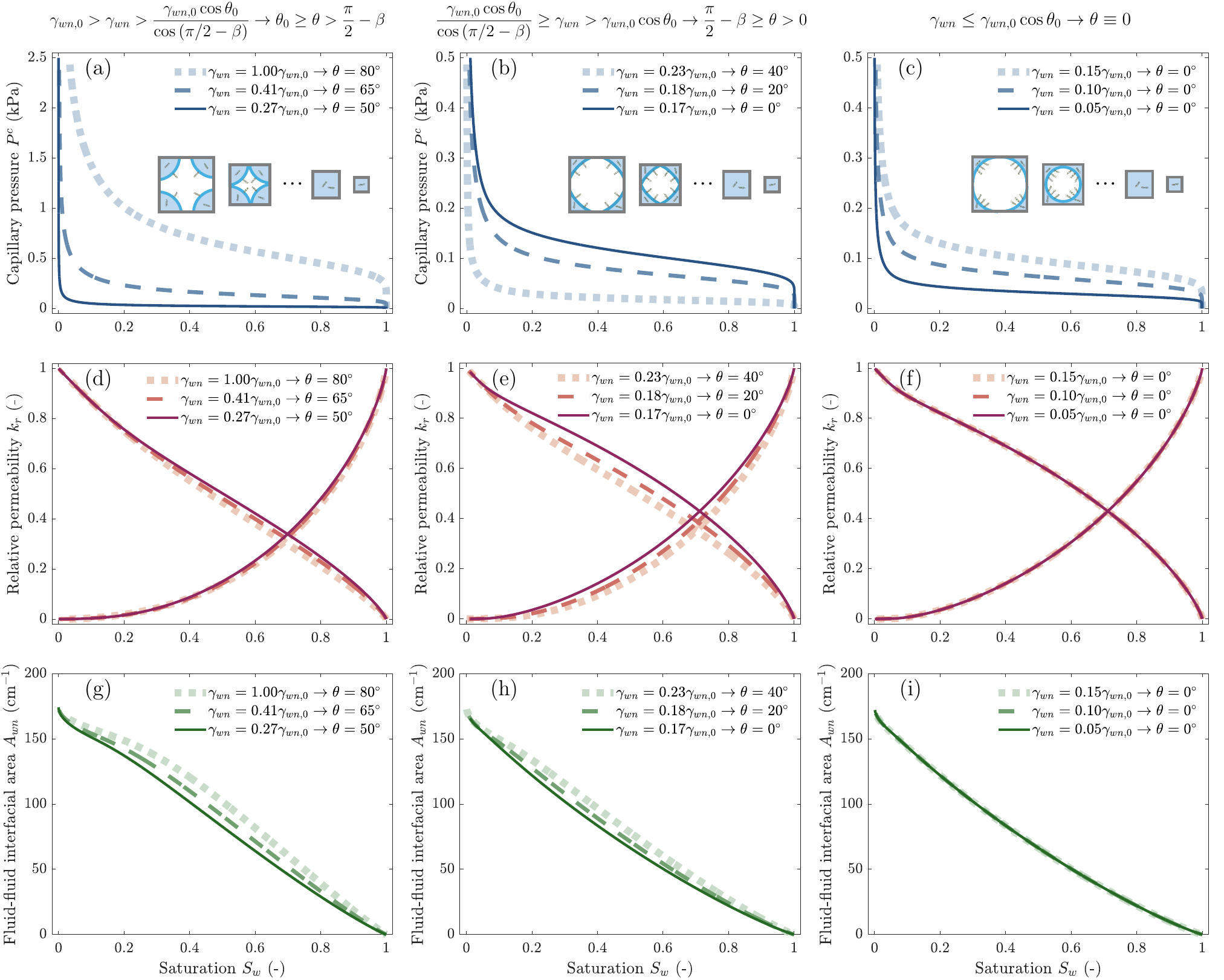}
  \caption{Impact of interfacial tension ($\gamma_{wn}$) on (a--c) capillary pressure--saturation curves, (e--f) relative permeability--saturation curves, and (g--i) fluid-fluid interfacial area--saturation curves in a porous medium with square-tube pores. A near neutral-wet condition (e.g., an intrinsic contact angle $\theta_0$ of $80^{\circ}$) is used as an illustrative example. Three $\gamma_{wn}$ ranges are examined: (1) A range within which the contact angles ($\theta$) are greater than 0 and are equal or greater than $\pi/2-\beta$, where $\beta$ is the half corner angle and $\beta = \pi/4$ in the square tubes (left column), (2) A range where $0< \theta < \pi/2-\beta$ (middle column), and (3) A range where $\theta \equiv 0$ (right column). In the figure, $\gamma_{wn,0}$ refers to the interfacial tension for surfactant-free fluids. Note that $\gamma_{wn}$ is a proxy of the surfactant effect, i.e., a smaller $\gamma_{wn}$ corresponds to a more interfacially active surfactant and/or a higher surfactant concentration.}
  \label{fig:squ-pore}
\end{figure}

\begin{figure}[!htb]
  \centering
  \includegraphics[width=1.0\textwidth]{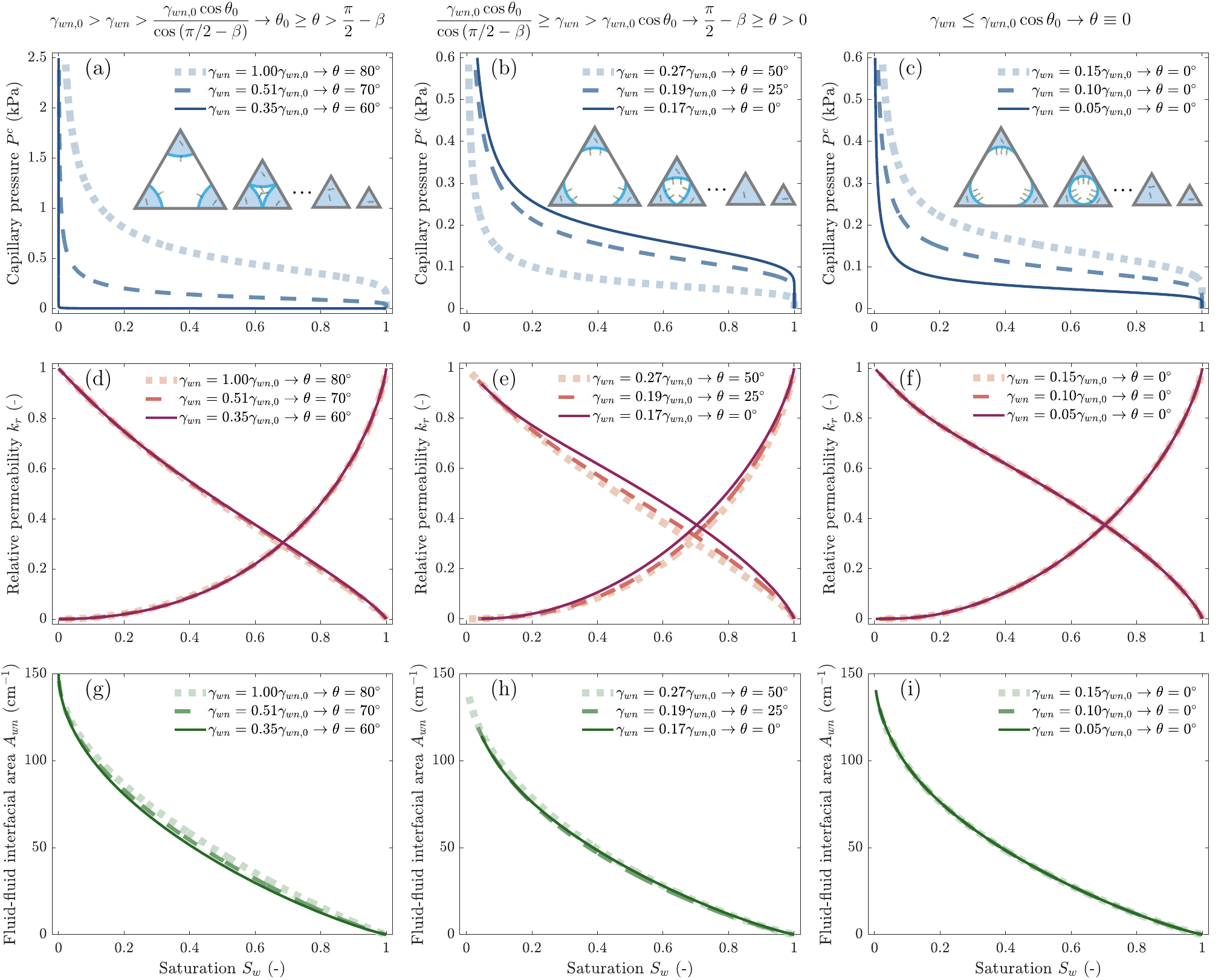}
  \caption{Impact of interfacial tension ($\gamma_{wn}$) on (a--c) capillary pressure--saturation curves, (e--f) relative permeability--saturation curves, and (g--i) fluid-fluid interfacial area--saturation curves in a porous medium with triangular-tube pores. A near neutral-wet condition (e.g., an intrinsic contact angle $\theta_0$ of $80^{\circ}$) is used as an illustrative example. Three $\gamma_{wn}$ ranges are examined: (1) A range within which the contact angles ($\theta$) are greater than 0 and are equal or greater than $\pi/2-\beta$, where $\beta$ is the half corner angle and $\beta = \pi/6$ in the equilateral triangular tubes (left column), (2) A range where $0< \theta < \pi/2-\beta$ (middle column), and (3) A range where $\theta \equiv 0$ (right column). In the figure, $\gamma_{wn,0}$ refers to the interfacial tension for surfactant-free fluids. Note that $\gamma_{wn}$ is a proxy of the surfactant effect, i.e., a smaller $\gamma_{wn}$ corresponds to a more interfacially active surfactant and/or a higher surfactant concentration.}
  \label{fig:tri-pore}
\end{figure}

In porous media containing square and equilateral triangular tubes, the impact of $\gamma_{wn}$ on two-phase flow properties becomes nonmonotonic and nonlinear. In particular, the impact of $\gamma_{wn}$ on the $P^c$--$S_w$ curves can be characterized by the following three distinct regimes:

\begin{itemize}
    \item Regime I: $\gamma_{wn} > \gamma_{wn,0} \cos \theta_0/\cos(\pi/2-\beta)$, corresponding to less interfacially active surfactants and/or lower surfactant concentrations. In this regime, the fluid menisci are convex and the radius of meniscus curvature ($r$) increases with the decrease of $\gamma_{wn}$. Since $P^c$ scales with $\gamma/r$, the decreasing $\gamma_{wn}$ and increasing $r$ lead to a decrease of $P^c$. Consequently, the $P^c$--$S_w$ curve shifts {\it downwards} with the decrease of $\gamma_{wn}$, i.e., the increase of surfactant concentration (Figures \ref{fig:squ-pore}a and \ref{fig:tri-pore}a).

    \item Regime II: $\gamma_{wn,0} \cos \theta_0 < \gamma_{wn} \le \gamma_{wn,0} \cos \theta_0/\cos(\pi/2-\beta)$, corresponding to moderately interfacially active surfactants and/or intermediate surfactant concentrations. In this regime,  the menisci become concave and $r$ begins to decrease with $\gamma_{wn}$ (Figures \ref{fig:squ-pore}b and \ref{fig:tri-pore}b). Because $P^c \propto \gamma_{wn} / r$ and $\gamma_{wn}$ decreases much less significant than $r$, the decrease of $\gamma_{wn}$ will increase the $P^c$ at the same $S_w$ and thus shift the $P^c$--$S_w$ curve {\it upwards} (Figures \ref{fig:squ-pore}b and \ref{fig:tri-pore}b). 

    \item Regime III: $\gamma_{wn} \le \gamma_{wn,0} \cos \theta_0$, corresponding to more interfacially active surfactants and/or higher surfactant concentrations. The $P^c$--$S_w$ curve switches back to a {\it downwards} shifting mode with the further decrease of $\gamma_{wn}$ given that $\theta \equiv 0$ and thus $P^c$--$S_w$ curves {\it scale linearly} with $\gamma_{wn}$ (Figures \ref{fig:squ-pore}c and \ref{fig:tri-pore}c). 
\end{itemize}
In contrast to $P^c$--$S_w$ curves, $k_r$--$S_w$ and $A_{wn}$--$S_w$ curves present much less significant variations with $\gamma_{wn}$ (Figure \ref{fig:squ-pore}d--i). This is because $k_r$ and $A_{wn}$ are mainly controlled by the fluid menisci sizes (i.e., the fluid perimeter and/or area in the cross-section of each unsaturated pore), which are much less sensitive to the $\gamma_{wn}$-dependent convex vs. concave shapes of fluid menisci than $r$ (Equations (\ref{eq:kw}--\ref{eq:knw}) and (\ref{eq:awn})).

The $\gamma_{wn}$-dependent two-phase flow properties can strongly influence two-phase flow and surfactant transport behaviors in porous media with varying pore geometries. To capture these effects quantitatively, we incorporate the new properties into Darcy-scale transient two-phase flow and surfactant transport models (Section~\ref{subsec:DarcyModel}).

\subsection{Validity and accuracy of explicit and closed-form expressions for two-phase flow properties} \label{subsec:ExplicitEXPvsClosedForm}

\subsubsection{Numerical experiment design}
To couple the $\gamma_{wn}$-dependent two-phase flow properties into Darcy-scale models, we have derived their explicit and closed-form expressions in Section~\ref{subsec:expression-function}. In this section, we evaluate the validity and accuracy of the explicit and closed-form expressions to ensure their reliability. This is achieved by comparing their predictions with those from the bundle-of-capillary-tubes model. Comparisons are performed for the three pore geometries at three representative surfactant concentrations (i.e., $\gamma_{wn}$ values). Similarly, $\gamma_{wn}$ is used as a proxy for the surfactant concentration---a larger $\gamma_{wn}$ corresponds to a lower surfactant concentration. Here, the $\gamma_{wn}$ values are chosen to capture the aforementioned two or three different regimes where the dependence of two-phase flow properties on pore annularity and $\gamma_{wn}$ exhibits distinct behaviors (Section~\ref{subsec:Impact2Properties}). For cylindrical pores, we use $\gamma/\gamma_{wn,0} = 0.22$, $0.10$, and $0.05$, corresponding to $\theta = 40^{\circ}$, $0^{\circ}$, and $0^{\circ}$. For square-tube pores, we use $\gamma/\gamma_{wn,0} = 0.41$, $0.18$, and $0.05$, corresponding to $\theta = 65^{\circ}$, $20^{\circ}$, and $0^{\circ}$. For equilateral triangular pores, we use $\gamma/\gamma_{wn,0} = 0.51$, $0.25$, and $0.05$, corresponding to $\theta = 70^{\circ}$, $45^{\circ}$, and $0^{\circ}$ (Note that angles are expressed in degrees ($^{\circ}$) hereafter, unless otherwise specified). Two pore size distributions are examined: a more uniform pore size distribution where the mean and normalized standard deviation of the pore sizes are respectively $100\;{\rm \mu m}$ and $0.3$ (i.e., the same as that in Section~\ref{subsec:Impact2Properties}), and a wider pore size distribution where the mean and normalized standard deviation of the pore sizes are respectively $100\;{\rm \mu m}$ and $0.5$ (Figure \ref{fig:psd}b).

\begin{figure}[!htb]
  \centering
  \includegraphics[width=1.0\textwidth]{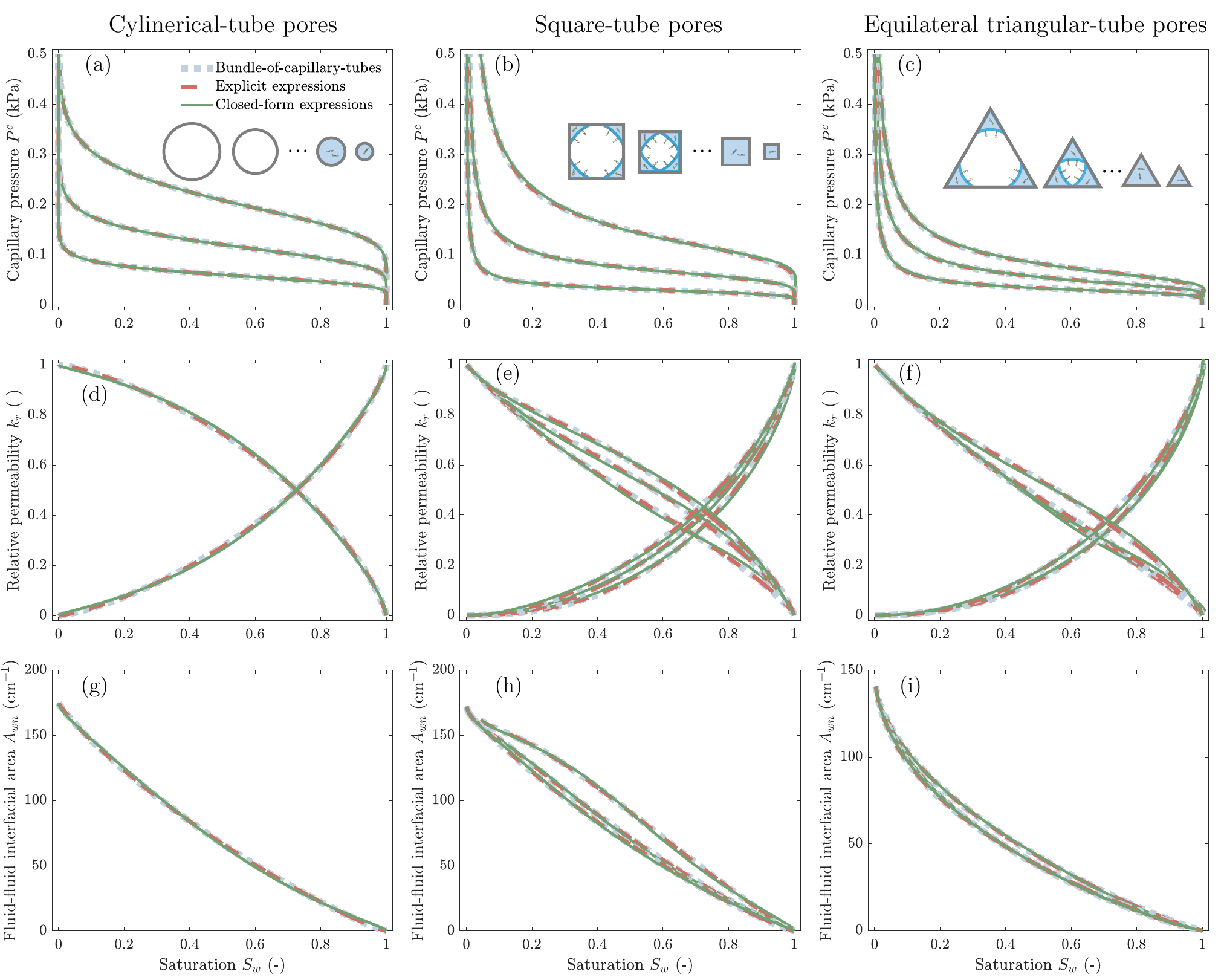}
  \caption{Evaluations on the explicit and closed-form expressions for (a--c) capillary pressure--saturation curves, (e--f) relative permeability--saturation curves, and (g--h) fluid-fluid interfacial area--saturation curves using the bundle-of-capillary-tubes model as a benchmark. Porous media with cylindrical, square-tube, and equilateral triangular-tube pores are examined. A near neutral-wet condition (e.g., an intrinsic contact angle of $80^{\circ}$) is used as an example. We present the curves at three example interfacial tensions ($\gamma_{wn}$) for each pore geometry---i.e., $\gamma_{wn}/\gamma_{wn,0} = 0.22$, $0.1$, and $0.05$ (i.e., $\theta = 40^{\circ}$, $0^{\circ}$ and $0^{\circ}$) for cylindrical pores; $\gamma_{wn}/\gamma_{wn,0} = 0.41$, $0.18$, and $0.05$ (i.e., $\theta = 65^{\circ}$, $20^{\circ}$ and $0^{\circ}$) for square-tube pores; and $\gamma_{wn}/\gamma_{wn,0} = 0.51$, $0.25$, and $0.05$ (i.e., $\theta = 70^{\circ}$, $45^{\circ}$ and $0^{\circ}$) for equilateral triangular-tube pores. The pore sizes follow a {\it full} lognormal distribution. The mean and normalized standard deviation of the pore sizes are respectively $100\;{\rm \mu m}$ and $0.3$.}
  \label{fig:explict-exp-closed-form-0.3}
\end{figure}

\subsubsection{Full lognormal pore size distribution}

As shown in Figure \ref{fig:explict-exp-closed-form-0.3}, the explicit expressions fully overlap with the results from the bundle-of-capillary-tubes model in all the cases, which verifies the mathematical consistency between the explicit expressions and the bundle-of-capillary-tubes model. Similarly, the closed-form expressions---which use a logistic function (i.e., $1/[1 + \exp(-1.702x)]$) to approximate the cumulative normal distribution function (i.e., $\Phi(x)$) in the explicit expressions---align excellently with the bundle-of-capillary-tubes model. We observe a slight deviation in the fluid--fluid interfacial area--saturation ($A_{wn}$--$S_w$) curves at $S_w \rightarrow 1$. At $S_w \rightarrow 1$, most pores are unsaturated. The $A_{wn}$ approximately scale with $1 - \Phi(x)$ where $x = (\ln R_{\rm max}-\mu)/\sigma$ and $R_{\rm max} \rightarrow \infty$ with $R_{\rm max}$ being the maximum saturated pore size (Equations \ref{eq:kr2-sw-anal}--\ref{eq:A12-angular-sw-anal}). Because the logistic function has a larger error at $x \rightarrow \infty$, $k_{r,w}$ and $A_{wn}$ present a larger error at $R_{\rm max} \rightarrow \infty$ (i.e., $S_w \rightarrow 1$). In contrast, $A_{wn}$ is near zero, and the corresponding fluid--fluid interfacial adsorption of surfactant is expected to be minor.

When the standard deviation of pore sizes increases to 0.5, both the explicit and closed-form expressions notably deviate from the bundle-of-capillary-tubes model (Figure \ref{fig:explict-exp-closed-form-0.5-log-normal}). This discrepancy is because the explicit and closed-form expressions cover the full range of a lognormal pore size distribution, while the bundle-of-capillary-tubes model can only approximate the lognormal pore size distribution by a truncated lognormal distribution.

\begin{figure}[!htb]
  \centering
  \includegraphics[width=1.0\textwidth]{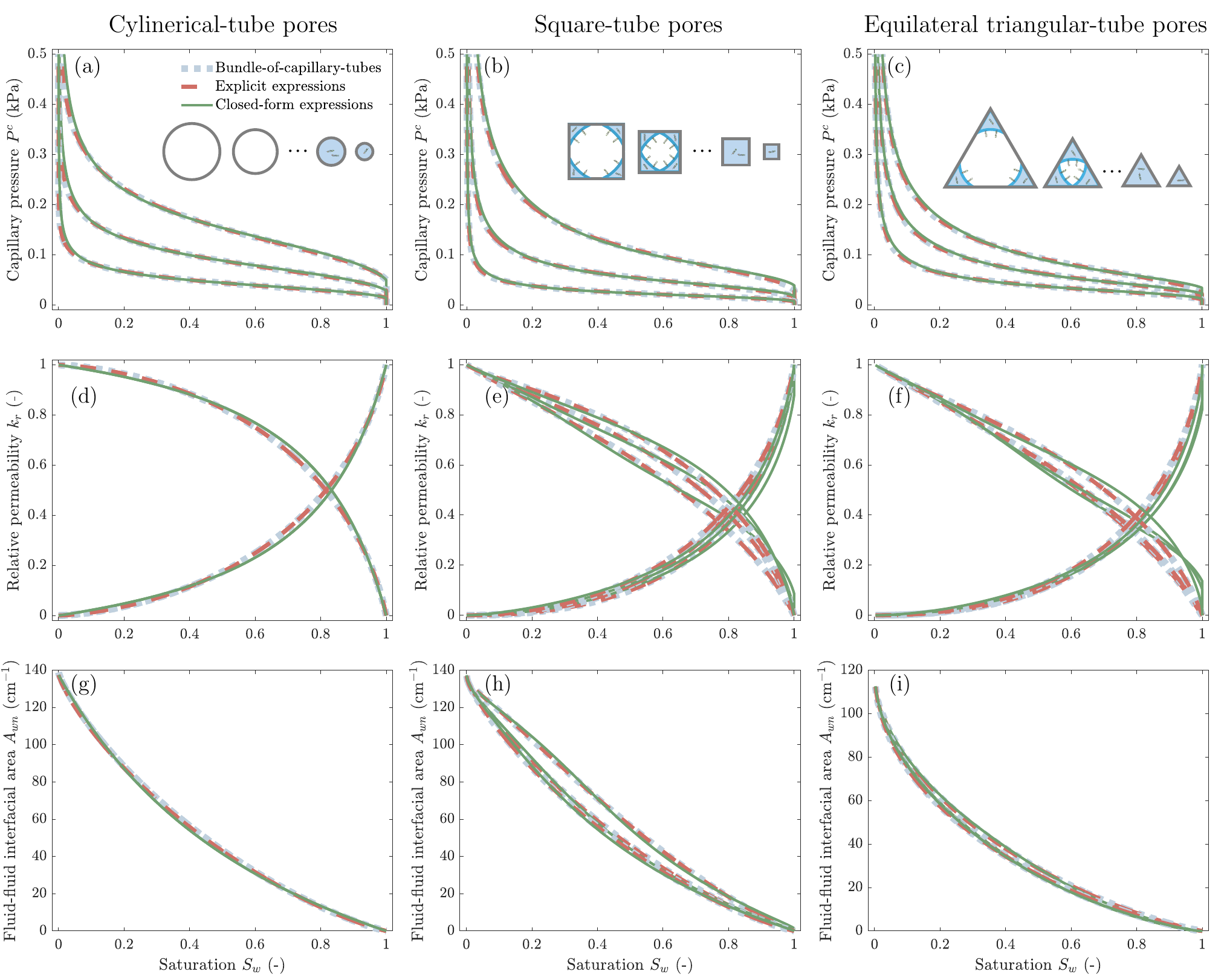}
  \caption{Evaluations on the explicit and closed-form expressions for (a--c) capillary pressure--saturation curves, (e--f) relative permeability--saturation curves, and (g--h) fluid-fluid interfacial area--saturation curves using the bundle-of-capillary-tubes model as a benchmark. Porous media with cylindrical, square-tube, and equilateral triangular-tube pores are examined. A near neutral-wet condition (e.g., an intrinsic contact angle of $80^{\circ}$) is used as an example. We present the curves at three example interfacial tensions ($\gamma_{wn}$) for each pore geometry---i.e., $\gamma_{wn}/\gamma_{wn,0} = 0.22$, $0.1$, and $0.05$ (i.e., $\theta = 40^{\circ}$, $0^{\circ}$ and $0^{\circ}$) for cylindrical pores; $\gamma_{wn}/\gamma_{wn,0} = 0.41$, $0.18$, and $0.05$ (i.e., $\theta = 65^{\circ}$, $20^{\circ}$ and $0^{\circ}$) for square-tube pores; and $\gamma_{wn}/\gamma_{wn,0} = 0.51$, $0.25$, and $0.05$ (i.e., $\theta = 70^{\circ}$, $45^{\circ}$ and $0^{\circ}$) for equilateral triangular-tube pores. The pore sizes follow a {\it truncated} lognormal distribution. The mean and normalized standard deviation of the pore sizes are $100\;{\rm \mu m}$ and $0.5$, respectively; while the minimum and maximum pore sizes are $5\;{\rm \mu m}$ and $500\;{\rm \mu m}$, respectively.}
  \label{fig:explict-exp-closed-form-0.5}
\end{figure}

\subsubsection{Truncated lognormal pore size distribution}
To address the above discrepancy, we use a truncated lognormal pore size distribution (where the minimum and maximum pore sizes are $5\;{\rm \mu m}$ and $500\;{\rm \mu m}$, respectively) to derive the explicit and closed-form expressions following the same procedure in Section \ref{subsec:expression-function}. As shown in Figure \ref{fig:explict-exp-closed-form-0.5}, the explicit expressions overlap with the bundle-of-capillary-tubes model that applies the same truncated lognormal pore size distribution, verifying the mathematical consistency and accuracy between the two models. The closed-form expressions have minor errors in calculating the capillary pressure and fluid--fluid interfacial area--saturation curves. However, the calculated relative permeability--saturation curves notably deviate from the reference curves, especially at $S_w \rightarrow 1$ where only a smaller amount of large pores are invaded by the nonwetting-phase fluid. The reason for this divination is the following. At $\sigma = 0.5$, the proportion of larger pores is greater than that at $\sigma = 0.3$. The invaded larger pores control the nonwetting-phase permeability and contribute a significant amount of wetting-phase permeability ($k_{nw} \propto R^4$). Additionally, the logistic function (i.e., $1/[1 + \exp(-1.702x)]$) makes larger errors at smaller or larger $x$. Collectively, the greater number of larger pores and a more significant error of the logistic function lead to a notable deviation.

Consequently, closed-formed functions are suggested for the capillary pressure--saturation curves of the two-phase flow models, while explicit expressions are recommended for the relative permeability--saturation curves. Either closed-form expressions or explicit expressions can be employed for the interfacial area--saturation curves of the surfactant transport models, since the interfacial area only enters the transport equation through surfactant mass balance and does not critically affect model structure.

\subsection{Modeling coupled unsaturated water flow and PFAS transport in soils}\label{subsec:model-couple}

The explicit and closed-form expressions are coupled into a Darcy-scale two-phase flow and surfactant transport (Section~\ref{subsec:DarcyModel}). Furthermore, the model is used to examine the impact of pore angularity and surfactant concentration on the coupled transient flow and transport processes for an example application, i.e., PFAS transport and leaching in soils. In particular, we simulate unsaturated water flow and PFAS transport through a vertical homogeneous soil column, which resembles a miscible displacement laboratory experiment (i.e., a laboratory experiment in which water containing a dissolved solute is injected into a soil column under constant infiltration conditions). The model employs the closed-form expressions for the capillary pressure--saturation ($P^c$--$S_w$) curves, and the explicit expressions for the relative permeability ($k_r$) and fluid--fluid interfacial area--saturation ($A_{wn}$--$S_w$) curves. The details in the modeling setup and simulation results are presented below.

\subsubsection{Numerical experiment design}

\begin{figure}[!htb]
  \centering
  \includegraphics[width=1.0\textwidth]{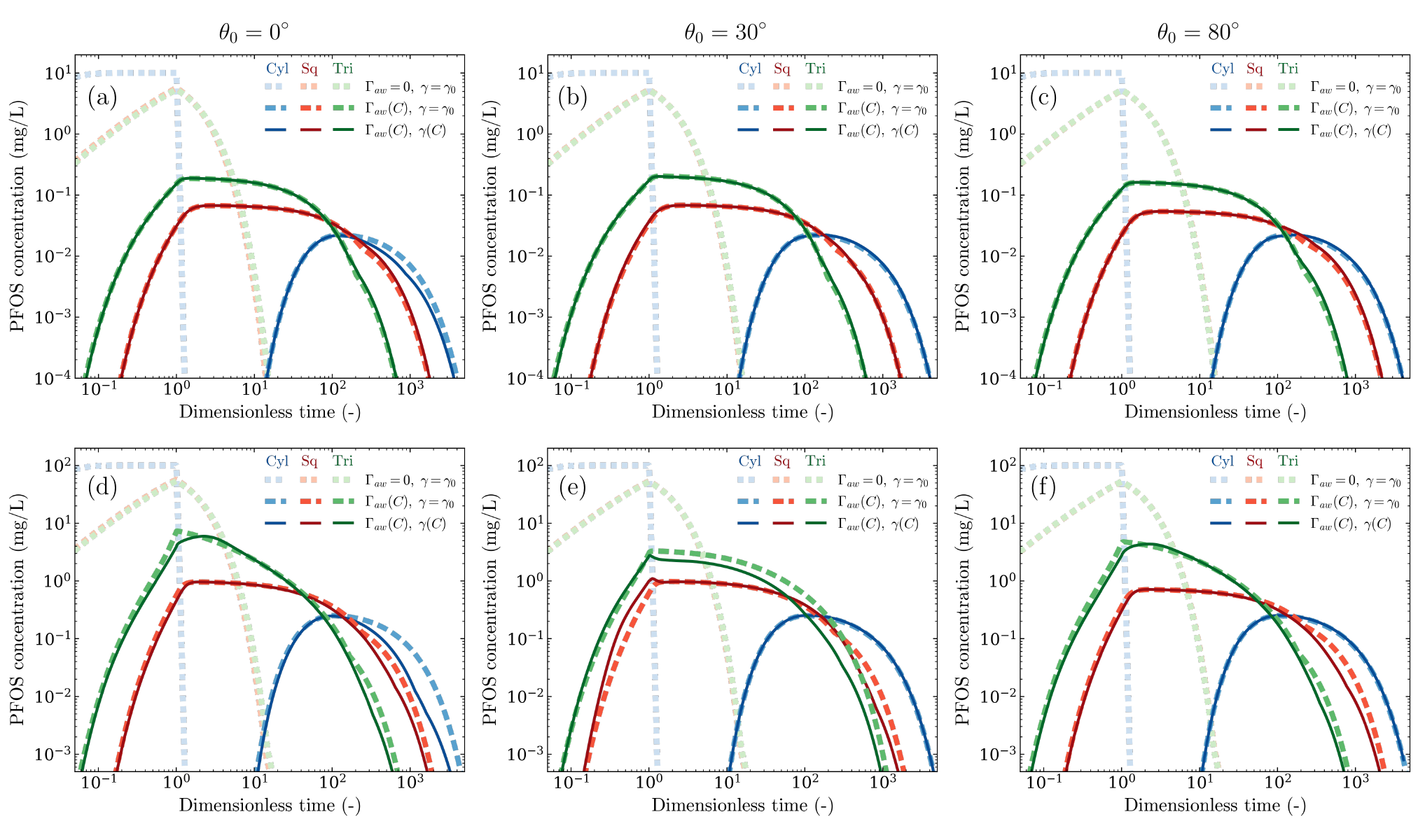}
  \caption{Impact of pore annularity, PFOS concentration, and intrinsic surface wettability on PFOS breakthrough curves in an unsaturated soil column. We consider two surface wettability: (a) completely water-wet ($\theta_0 = 0^{\circ}$), (b) partially water-wet ($\theta_0 = 30^{\circ}$), and (c) weakly water-wet ($\theta_0 = 80^{\circ}$). For each surface wettability, we compare the PFOS breakthrough curves for three pore geometries: cylindrical tubes (denoted by ``Cyl''), square tubes (denoted by ``Sq''), and triangular tubes (denoted by ``Tri''). The breakthrough curves are plotted on a log scale for both $x$ and $y$ axes. On the $x$-axis, the time is nondimensionalized by $L/q$, where $L$ is the length of the soil column and $q$ is the infiltration rate.}
  \label{fig:PFAS-btc}
\end{figure}

We simulate the unsaturated water flow and PFAS transport behavior in the soil column using three versions of the model: (1) base case, which turns off the fluid--fluid interfacial adsorption of PFAS and its impact on interfacial tension (i.e., $\Gamma_{aw} = 0$ and $\gamma \equiv \gamma_{0}$, where $\Gamma_{aw}$ is the air--water interfacial excess, $\gamma$ is the surface tension for PFAS-laden water, $\gamma_{0}$ is the surface tension for PFAS-free water); (2) intermediate case, which turns on the fluid--fluid interfacial adsorption while keeping a constant $\gamma$ (i.e., i.e., $\Gamma_{aw} = K_{aw} C$ and $\gamma \equiv \gamma_{0}$); (3) full model, which accounts for both the fluid--fluid interfacial adsorption of PFAS and its impact on $\gamma$ (i.e., i.e., $\Gamma_{aw} = K_{aw} C$ and $\gamma = \gamma_{0} \left[1 - b\ln(1 + C/a)\right]$). Each model is applied to porous media with different pore geometries (cylindrical, square-tube, and equilateral triangular-tube) and intrinsic contact angles ($\theta_0=0^{\circ}$, $30^{\circ}$, and $80^{\circ}$). We use PFOS as an example PFAS substance and simulate two representative aqueous concentrations (10\,mg/L and 100\,mg/L), yielding 18 total simulations.

All simulations begin with a constant infiltration of PFOS-free water into a 10\,cm-long dry soil column from the inlet (top), while the outlet (bottom) is maintained under free drainage. A small water saturation (0.001) is used to initiate the infiltration simulation. After the infiltration reaches the steady state, a pulse of PFAS solution (two pore volumes) is introduced at the same infiltration rate, with a zero concentration gradient imposed at the outlet. The parameters used for water flow and PFOS transport are summarized in the following: the infiltration rate is 0.42\,cm/h, $\phi = 0.395$\,cm$^3$/cm$^3$, $k_0 = 1.19\times10^{-8}$\,cm$^2$, $D_0 = 5.4\times 10^{-6}$\,cm$^2$/s, $\mathcal{L}_w = 34.96$\,cm, $\gamma_0= 0.072$\,N/m, $R_g = 8.314$\,J/K/mol, $T = 293.15$\,K, $a$ = 4$\times10^{-3}\,$mol/m$^3$, $b$ = 0.107.

\subsubsection{Unsaturated water flow and PFAS transport in soils}

For each simulation, we compute the breakthrough curves at the outlet (Figure \ref{fig:PFAS-btc}). As expected, PFOS shows significant retention due to adsorption at the air–water interface. However, the extent of retention varies strongly with pore geometry. The retardation factors (defined as the ratio of solute transport velocity to pore-water velocity) are approximately 306.5 for cylindrical pores, 6.6 for square-tube pores, and 3.0 for triangular-tube pores. This variation arises because pore geometry controls relative permeability and, in turn, water saturation under constant infiltration. Under steady infiltration of surfactant-free water, pore angularity alone leads to an order-of-magnitude difference in water saturation. As a result, the air–water interfacial areas differ markedly across geometries (134.4\,cm$^{-1}$ for cylindrical pores, 48.6\,cm$^{-1}$ for square-tube pores, and 24.8\,cm$^{-1}$ for triangular-tube pores), which directly translates into an order-of-magnitude difference in retardation factors. These results highlight the critical role of pore geometry in shaping two-phase flow properties---particularly relative permeability and air–water interfacial area---and thereby governing the transport of surface-active solutes such as PFAS.

As shown in Figure \ref{fig:PFAS-btc}a--c, the models with and without accounting for interfacial tension variations produce nearly identical breakthrough curves for all three pore geometries at the lower concentration (10\,mg/L), where the interfacial tension remains essentially unchanged. At the higher concentration (100\,mg/L), the interfacial tension variations lead to a slight surfactant-induced flow and a slightly weaker retention (Figure \ref{fig:PFAS-btc}d--f).

\begin{figure}[!htb]
  \centering
  \includegraphics[width=1.0\textwidth]{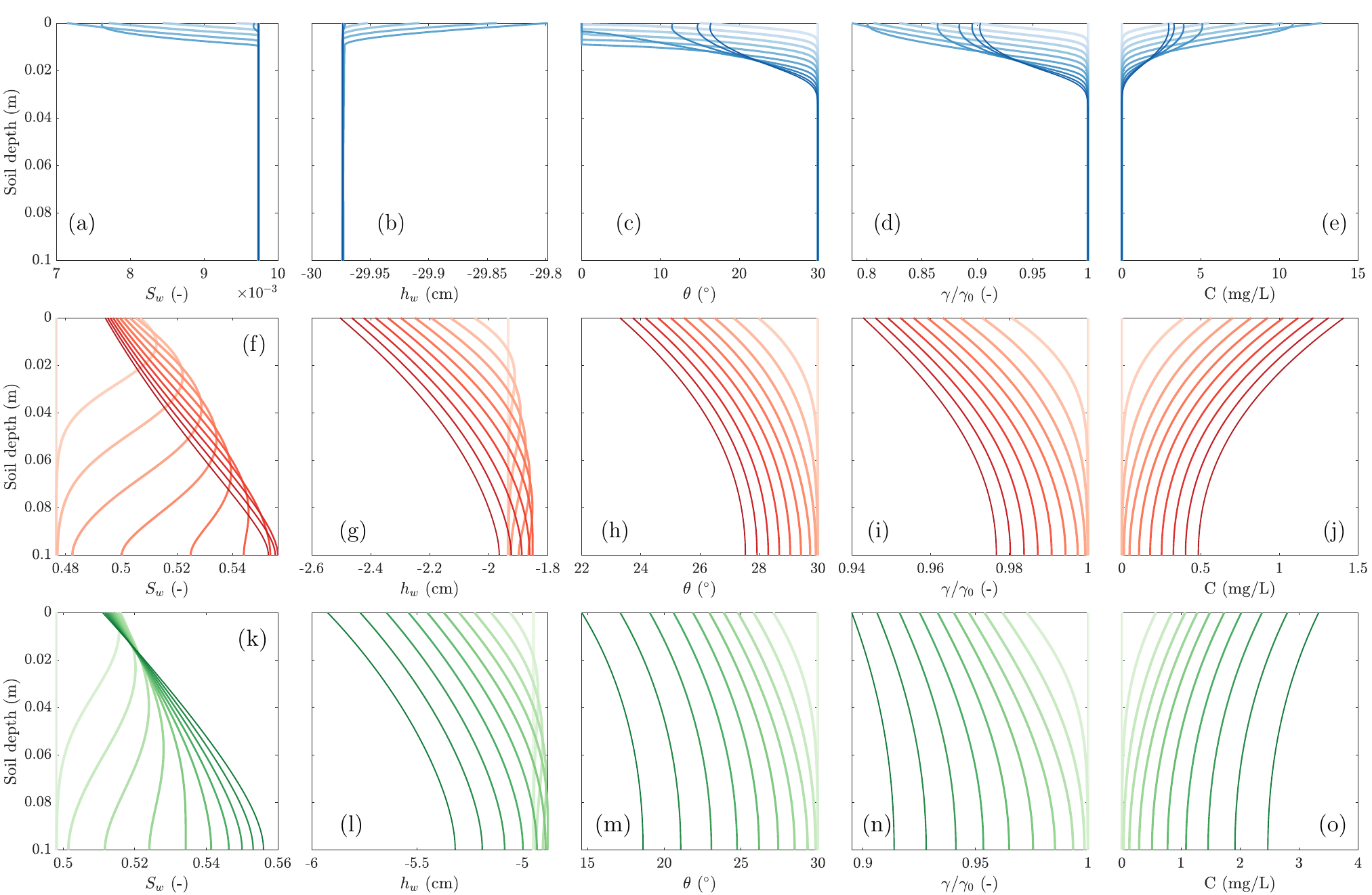}
  \caption{Temporal and spatial variations of water saturation ($S_w$), water pressure head ($h_w$), contact angle ($\theta$), interfacial tension ($\gamma$, which is normalized by the interfacial tension for pure water $\gamma_0$), and aqueous PFOS concentration ($C_{aq}$) in an unsaturated soil column. We present the simulation results for all three pore geometries---including cylindrical tubes (top row), square tubes (middle row), and triangular tubes (bottom row)---at the PFOS concentration of 100\,mg$/$L under partially water-wet condition ($\theta_0 = 30^{\circ}$). We present the profiles from 0 to 4\,hours at an interval of 0.4\,hour for the porous medium with cylindrical pores, and those from 0 to 2\,hours at an interval of 0.2\,hour for the porous medium with angular pores. The lines become progressively darker to represent advancing simulation time.}
  \label{fig:PFAS-profile-partial-wet}
\end{figure}

A close inspection of the soil column with cylindrical-tube pores and a contact angle of $30^{\circ}$ shows that both interfacial tension and contact angle vary substantially in space and time. For an intrinsic contact angle of $30^{\circ}$, the simulations show that the interfacial tension decreases by up to 20\% and the contact angle by as much as $30^{\circ}$ (Figure \ref{fig:PFAS-profile-partial-wet}c--d). These changes lead to a 30\% reduction in water saturation in the upper soil and a 0.7\% shift in water pressure head therein (Figure \ref{fig:PFAS-profile-partial-wet}a--b). The fluctuations occur mainly in the top soils due to the strong retention and the resulting slower downward migration and spreading of PFOS across the soil column. Nevertheless, the greater relative permeability of cylindrical pores results in a lower water saturation and a larger air–water interfacial area at the same infiltration rate. This enlarged interfacial area promotes strong PFOS adsorption (Figure \ref{fig:PFAS-profile-partial-wet}e), which in turn damps the fluctuations in water saturation and PFOS concentration induced by surfactant-driven flow. Therefore, the breakthrough curves only show a minor difference even at the higher concentration.



In soil columns with angular pores and an intrinsic contact angle of $30^{\circ}$ (Figure \ref{fig:PFAS-profile-partial-wet}h–i, m–n), the reductions in interfacial tension ($<$10\%) and contact angle ($<15^{\circ}$) are much smaller. Correspondingly, water saturation varies by under 12\% (Figure \ref{fig:PFAS-profile-partial-wet}f\&k). However, the water pressure head changes are more pronounced (about 44\% for square-tube pores and 20\% for triangular-tube pores) due to the weaker retention and thus faster downward migration and spreading of PFOS in the column (Figure \ref{fig:PFAS-profile-partial-wet}g\&l). To assess the role of the water pressure head changes, we compared the water pressure head gradient with the gravitational force gradient across the entire domain. The computed gradients range from 0.08 to 0.12, which is much smaller than the gravitational gradient ($\partial h/\partial z = 1$, where $h$ is the hydraulic head). Thus, the spatial variations in PFOS concentration, interfacial tension, and contact angle exert little influence on PFOS transport under these simulated conditions (Figure \ref{fig:PFAS-profile-partial-wet}j\&o).

In soils with intrinsic contact angles of $0^{\circ}$ and $80^{\circ}$, the temporal and spatial variations of water saturation, pressure head, and contact angle slightly deviate from each other (Figures \ref{fig:PFAS-profile-comlete-wet} and \ref{fig:PFAS-profile-weak-wet}), due to the distinct dependence of capillary pressure--saturation curves on interfacial tension at different intrinsic contact angles. However, because of the insignificant surfactant-induced flow and strong retention, the PFOS concentration profiles present similar patterns as those in soils with an intrinsic contact angle of $30^{\circ}$ (Figures \ref{fig:PFAS-profile-comlete-wet} and \ref{fig:PFAS-profile-weak-wet}). Correspondingly, the breakthrough curves are similar for different intrinsic contact angles, even at a higher concentration (100\,mg/L).

\section{Discussion}
\label{sec:discussion}

We present an upscaling workflow for simulating the coupled two-phase surfactant/PFAS-laden fluid flow and surfactant/PFAS transport in porous media with angular pores. The workflow derives REV-scale two-phase flow properties directly from pore-scale structural information (including intrinsic surface wettability, pore geometry, and pore-size distribution) and surfactant chemistry conditions. These REV-scale properties are then coupled into a transient two-phase flow and surfactant/PFAS transport model, enabling prediction of complex flow and transport behaviors across a wide range of porous media using only a minimal set of pore-scale structural parameters. The approach offers a practical tool for applications such as contaminant transport and remediation (e.g., NAPL and PFAS) in soils and groundwater \citep[e.g.,][]{al2009impact,guo2020mathematical,maroli2024surfactant}.

\subsection{New scaling functions for two-phase flow properties of surfactant-laden fluids in porous media}\label{subsec:new-scaling}

Through modeling investigations, we identify new scaling functions for the two-phase flow properties of surfactant-laden fluids in porous media, distinct from the classic Leverett $J$-function \citep{leverett1941capillary} and its modified forms \cite[e.g.,][]{rose1949evaluation,lan2025scaling}. The Leverett $J$-type scalings extrapolate the capillary pressure–saturation ($P^c$--$S_w$) curve of a given rock and surfactant-free fluid system to other rocks or fluid pairs by normalizing with $1/(\gamma_{wn,0}\cos\theta_0)$ or $1/(\gamma_{wn,0}f(\theta_0))$, where $\gamma_{wn,0}$ is the interfacial tension for surfactant-free fluids, $\theta_0$ is the intrinsic contact angle, and $f$ is a function of $\theta_0$. For surfactant-laden fluids, a common empirical extension is to substitute $\gamma_{wn,0}$ and $\theta_0$ with $\gamma_{wn}$ and $\theta$, where $\gamma_{wn}$ and $\theta$ are treated as independent empirical or semi-empirical functions of surfactant type and concentration \cite[e.g.,][]{bhattacharjee2025evolution}. However, as implied by the Young–Dupré equation, $\gamma_{wn}$ and $\theta$ are coupled. Our scaling functions, derived mechanistically from the Young–Dupré relation and the bundle-of-capillary-tubes model, take into account this coupling effect. They yield two-phase flow properties only depending on $\gamma_{wn}$ (i.e., surfactant types and concentrations) and pore structures.

Additionally, most previous studies adopted a linear Leverett $J$-type scaling of the $P^c$--$S_w$ curve and assumed invariant $K_r$--$S_w$ and $A_{wn}$--$S_w$ curves. In contrast, our scaling functions predict a piecewise linear relationship between $P^c$--$S_w$ and $\gamma_{wn}$ in porous media with cylindrical pores, and a monotonic, nonlinear relationship in porous media with angular pores. Furthermore, the relative permeability--saturation ($K_r$--$S_w$) and fluid–fluid interfacial area--saturation ($A_{wn}$--$S_w$) curves both vary with $\gamma_{wn}$. The new scaling functions may capture more realistic two-phase flow behaviors of surfactant-laden fluids in porous media with diverse geometries. Incorporating these new properties into Darcy-scale models is essential for accurately predicting transient two-phase flow and contaminant transport in porous media, particularly under conditions where surfactant effects are significant.

\subsection{Experimental validation of the proposed scaling functions}\label{subsec:expt-validation}

To validate the proposed scaling functions, we compare the model predictions against experimentally measured capillary pressure--saturation curves of surfactant(Triton X-100)-laden fluids in an F-70 Ottawa sand \citep{karagunduz2001influence}. The comparison involves two steps: (1) We compute the pore size distribution using the van Genuchten (VG) curve fitting best with the experimentally measured capillary pressure--saturation curve of surfactant-free fluids. Specifically, we can compute the cumulative density function of the pore sizes via $F_{\rm V}(R_{\rm max}) = S_{w,{\rm VG}}(P^c) = S_{w,{\rm VG}}(P^c(R_{\rm max}))$, where $F_{\rm V}$ is the cumulative volume of pores whose sizes are below $R_{\rm max}$, $S_{w,{\rm VG}}(P^c)$ is the VG model, $P^c(R_{\rm max})$ is given by the bundle-of-capillary-tubes model (Equation (\ref{eq:pc-sw})) or its simplified expressions with $\gamma = \gamma_0$ and $\theta = \theta_0$, and $R_{\rm max}$ is the maximum pore size below which the pores remain fully occupied by the wetting-phase fluid. Consequently, the probability density function of the pore sizes is given by $f_{\rm V}(R_{\rm max}) = \frac{{\rm d} F_{\rm V}(R_{\rm max})}{{\rm d} R_{\rm max}} = \frac{{\rm d} S_{w,{\rm VG}}(P^c)}{{\rm d} P^c} \frac{{\rm d} P^c (R_{\rm max})}{{\rm d} R_{\rm max}}$, where $f_{\rm V}$ represents the volume of pores with radius $R_{\rm max}$. We can divide $f_{\rm V}$ by the volume of a single pore to yield the corresponding pore count. (2) We predict the capillary pressure--saturation curves of surfactant-laden fluids using the derived pore size distribution and measured interfacial properties ($\gamma$ and $\theta$) on planar slides made of the same material as that of the studied porous medium, and compare the predictions with their corresponding experimentally measured capillary pressure--saturation curves. To minimize the occurrence and influence of more complex two-phase flow processes (e.g., trapping and hysteresis), the validation is done using experimental measurements during primary drainage. The specifics are provided below.

\begin{figure}[!htb]
  \centering
  \includegraphics[width=1.0\textwidth]{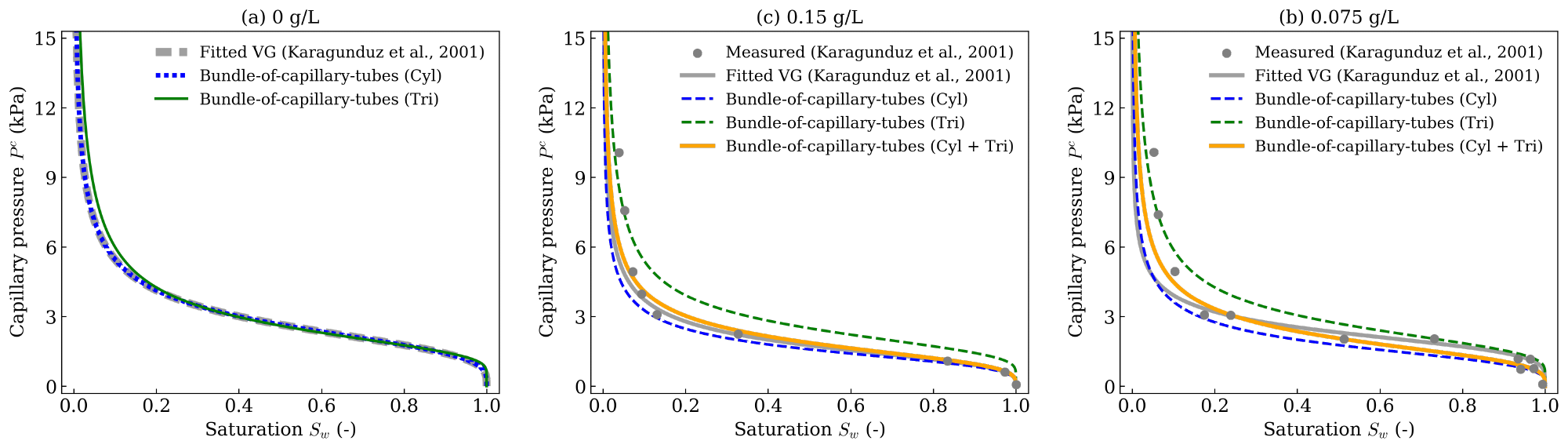}
  \caption{Experimental validation of the new scaling functions for two-phase flow properties of surfactant(Triton X-100)-laden fluids in an F-70 Ottawa sand \citep{karagunduz2001influence}. (a) Derivation of pore size distribution from the van Genuchten (VG) curve of surfactant-free fluids and validation of bundle-of-capillary-tubes models using cylindrical tubes (denoted by ``Cyl'') or triangular tubes (denoted by ``Tri''). (b--c) Validation of predicted capillary pressure--saturation curves for surfactant-laden fluids using the derived pore size distribution at two Triton X-100 concentrations: 0.075\,g/L and 0.15\,g/L. The predictions are provided by bundle-of-capillary-tubes models with three different pore geometries, including cylindrical tubes (denoted by ``Cyl''), triangular tubes (denoted by ``Tri''), and a combination of cylindrical and triangular tubes (denoted by ``Cyl\,+\,Tri'').  All VG curves were obtained by fitting with experimental data, and the VG parameters were reported by \citet{karagunduz2001influence}.}
  \label{fig:expt-validation}
\end{figure}

We use cylindrical and triangular tubes to approximate the rounded to sub-angular pore spaces in the F-70 Ottawa sand \citep[e.g.,][]{bastidas2016ottawa}. As shown in Figure \ref{fig:expt-validation}a, the bundle-of-capillary-tubes models using the VG curve-derived pore size distributions and measured interfacial properties ($\gamma_0$ and $\theta_0$, see Table \ref{tab:expt_measurement}) accurately reproduce the experimental data-fitted VG curve for surfactant-free fluids \citep{karagunduz2001influence}, verifying the theoretical derivation of pore size distributions. The same models are used to predict the capillary pressure--saturation curves of surfactant-laden fluids using measured interfacial properties ($\gamma$ and $\theta$, see Table \ref{tab:expt_measurement}) on quartz slides at two concentrations (i.e., 0.075\,g/L and 0.15\,g/L). We observe that the model using cylindrical tubes consistently overestimates the reduction of capillary pressure by surfactants, while the one using triangular tubes shows the opposite (Figure \ref{fig:expt-validation}b--c). These deviations may be caused by the coexistence of rounded and sub-angular pore spaces in the F-70 Ottawa sand. To account for this phenomena, we further estimate the curve via a weighted average of the two models, i.e., $S_w(P^c) = W \cdot S_{w, \rm Cyl}(P^c) + (1-W)\cdot S_{w, \rm Tri}(P^c)$ where $W$ is the volume fraction of cylindrical pores, and the subscripts ``Cyl'' and ``Tri'' represents the models using cylindrical and triangular tubes, respectively. Interestingly, the weighted average significantly improves the accuracy of the predictions at a wide range of $W$ (i.e., $W = 0.55 \sim 1.00$). For example, with $W = 0.65$, the root mean square errors between the predictions and measurements decrease from 0.128 (cylindrical tubes) and 0.104 (triangular tubes) to 0.079 (weighted average) at 0.075\,g/L, while decreasing from 0.040 (cylindrical tubes) and 0.126 (triangular tubes) to 0.028 (weighted average) at 0.15\,g/L. These errors are also close to the errors of the VG curves that were best fitted to the measurements (0.059 at 0.075\,g/L and 0.020 at 0.15\,g/L). Both the error reduction and the good match validate the predictive capability of the bundle-of-capillary-tubes models and the accuracy of the new scaling functions. 

More importantly, the above analyses suggest that, through an appropriate approximation of pore geometry and pore size distribution, the simplified bundle-of-capillary-tubes models can provide good predictions for the capillary pressure-saturation curves and potentially other two-phase flow properties of surfactant-laden fluids in real porous media using minimal experimental measurements (i.e., capillary pressure-saturation curves of surfactant-free fluids and interfacial properties of surfactant-laden fluids on planar slides of the same solid materials).

\subsection{Coupling of new two-phase flow properties into transient two-phase flow and surfactant transport model}\label{subsec:new-coupling}

We acknowledge that many prior studies have derived two-phase flow properties for surfactant-free fluids by accounting for factors such as thin fluid films, pore geometry, surface wettability, and surface roughness \citep[e.g.,][]{tuller1999adsorption,diamantopoulos2013physically,bhattacharjee2025evolution}, and have proposed explicit or closed-form expressions for these properties \citep[e.g.,][]{diamantopoulos2015closed}. However, to our knowledge, these developments have rarely been carried through into transient two-phase flow and transport models, even for surfactant-free fluid systems. This gap likely stems from the analytical complexity of the derived relationships, the absence of simplified forms suitable for numerical implementation, and/or the additional challenges of coupling them with nonlinear two-phase flow and transport processes. 

Here, we for the first time demonstrate the feasibility of coupling theoretically derived properties for surfactant-laden fluids and their simplified expressions into a transient two-phase flow and transport framework. The coupled model is solved using either a fully implicit or a sequentially implicit scheme, with preliminary tests suggesting that the sequentially implicit method offers superior numerical stability. While further investigation is needed to address the mathematical complexity and to develop more robust numerical algorithms, our example simulations of unsaturated water flow and PFAS (surfactant-like chemical) transport show that the model successfully captures both water flow fluctuations and PFAS transport dynamics in unsaturated soils. These findings demonstrate that the approach is not only theoretically sound but also practically feasible, paving the way for broader applications in coupled two-phase flow and surfactant transport systems

\subsection{Implications for PFAS transport in vadose zone soils}\label{subsec:pfas-transport}

Furthermore, the PFAS transport simulations show that PFAS migration and retention in unsaturated soils are strongly governed by pore geometry. Simulations under constant infiltration rates reveal sharp contrasts among pore types: cylindrical pores yield retardation factors on the order of 300, compared to 6.6 and 3.0 for square and triangular pores, respectively. This variation arises primarily from differences in water saturation and interfacial area under simulated infiltration conditions. Cylindrical pores maintain low saturation and large interfacial areas, conditions favorable for strong PFOS adsorption. In contrast, angular pores sustain higher saturation and smaller interfacial areas, leading to weaker retention. These findings suggest that soils dominated by cylindrical-like pores (if they exist) may act as long-term PFAS reservoirs, whereas soils with more angular pores may facilitate faster contaminant migration. This distinction highlights the importance of considering pore geometry in predicting PFAS transport and designing targeted remediation strategies.

In contrast to pore geometry, the simulations indicate that surfactant-induced flow has a subtle impact on PFAS (e.g., PFOS) transport. This finding is consistent with miscible displacement experiments conducted with both PFAS and non-PFAS surfactants at similar concentration levels \citep{brusseau2007measuring,brusseau2015novel,el2021testing}. At the field scale, however, the extent of surfactant-induced flow and its influence on PFAS transport remains unsettled. For example, \citet{zeng2021multidimensional} observed similar vertical concentration profiles between long-term one- and two-dimensional simulations for PFAS concentrations up to 1,000\,mg/L, concluding that surfactant-induced flow was negligible. In contrast, other studies suggest that surfactant-induced flow can play a more significant role in the lateral spreading of bulk aqueous fire-fighting foam at orders-of-magnitude higher concentrations, but without considering vertical flow and transport processes \citep{valvatne2005predictive}. These differences appear to depend strongly on PFAS type and concentration level, flow and transport representation (e.g., vertical vs.\ lateral vs.\ full two-/three-dimensional), as well as site-specific conditions such as infiltration rates and soil heterogeneity. To explore these factors, a two-/three-dimensional form of our model or other existing two-/three-dimensional models \citep[e.g.,][]{zeng2021multidimensional} can be applied. 

We recognize that the presented simulations focus on short-term PFAS transport within a shallow ($\sim$10\,cm) soil column. The results are intended to elucidate fundamental pore-scale mechanisms rather than to directly predict long-term field-scale behavior. The relevance of these mechanisms to deeper vadose-zone soils (>1\,m) and longer time scales---where PFAS leaching poses the greatest risk to groundwater---remains an open question. Addressing this gap will require future studies that link pore-scale retention and mass-transfer parameters to depth-resolved vadose-zone models, thereby enabling the evaluation of cumulative PFAS leaching and mass discharge under realistic meteorological and hydrologic forcing.

Finally, we note that while most prior PFAS transport studies applied a simplified two-phase flow model (i.e., Richards’ equation) to focus on transport in a single fluid such as water \citep[e.g.,][]{guo2020mathematical,silva2020modified}, our model---a generalization of Richards’ equation---accounts for full two-phase flow and PFAS transport in both air and water. This broader framework is particularly useful for representing systems where PFAS transport and mass transfer in both phases are significant. For example, volatile PFAS compounds require modeling transport in both air and water \citep[e.g.,][]{brusseau2024vapor}, and similar considerations apply to water–NAPL systems with substantial PFAS partitioning into both phases \citep[e.g.,][]{liao2022influence}. These advancements mark a critical step toward accurate long-term leaching predictions, site-specific risk assessments, and the design of effective remediation strategies in complex environmental settings.

\subsection{Model limitations and extensions}

While the model in this study adopts several simplifying assumptions---idealized pore geometry, lognormal or truncated lognormal pore size distribution, and uniform wettability, it can be extended to more complex conditions. Potential extensions include broader classes of angular pore geometries, multimodal pore size distributions, and mixed-wet porous media. For additional angular geometries beyond square and triangular tubes, the two-phase flow properties and their simplified mathematical forms can be derived using the methodology in Section \ref{sec:model}. In the case of a multimodal pore size distribution, the problem can be decomposed into a superposition of multiple single-modal distributions, with properties derived for each mode and then combined. For mixed-wet porous media, effective two-phase properties can be obtained by superposing contributions from pores of different intrinsic contact angles. These extensions will broaden the applicability of the framework to diverse surfactant-laden flow and porous media systems, including biosurfactant–water-soil environments \citep{yang2021effect,li2022corner}, oil and gas reservoirs \citep{lake1989enhanced}, and geological CO$_2$/H$_2$ storage formations \citep{foyen2020increased,chaturvedi2022air}.

The bundle-of-capillary-tubes model necessarily simplifies pore structure by neglecting spatial connectivity and arrangement. It may not capture some additional pore-scale phenomena (e.g., fluid trapping and hysteresis during cyclic drainage--imbibition experiments) under certain experimental and field conditions. When such processes are important, more realistic pore-scale models---such as pore morphology model \citep[e.g.,][]{hazlett1995simulation,hilpert2001pore}, pore-network model \citep[e.g.,][]{fatt1956network,reeves1996functional,chen2023pore}, and direct numerical simulations \citep[e.g.,][]{martys1996simulation,raeini2012modelling}---can be employed to derive two-phase properties. However, these approaches often do not yield closed-form analytical expressions, creating challenges for direct coupling with Darcy-scale models. In such cases, we can embed advanced pore-scale models (e.g., pore networks) directly within each grid block of a Darcy-scale model and form a multiscale framework \citep[e.g.,][]{blunt2002detailed}. Another practical and computationally efficient approach is employing empirical fitting to obtain tractable functions for Darcy-scale simulations. These developments hold promise for improving the predictive capability and accuracy of field-scale simulations under real-world conditions.

\section{Conclusion}
\label{sec:summary}

We present an upscaling workflow to quantify how pore-scale variations in surfactant/PFAS-associated interfacial properties (i.e., interfacial tension and contact angle) and pore angularity influence Darcy-scale two-phase flow and surfactant/PFAS transport. The workflow proceeds through four key steps: (1) Compute the fluid--fluid configuration in nonangular or angular pores for a given capillary pressure and interfacial tension (i.e., surfactant/PFAS type and concentration), (2) Determine fluid--fluid partitioning in a bundle of capillary tubes at the specified capillary pressure, (3) Derive two-phase flow properties (e.g., capillary pressure, relative permeability, and fluid--fluid interfacial area as functions of saturation) using the bundle-of-capillary-tubes model, (4) Couple these properties into Darcy-scale transient two-phase flow and surfactant/PFAS transport models. This pore-scale to Darcy-scale upscaling workflow provides a mechanistic tool to study the complex interactions between transient two-phase flow and surfactant/PFAS transport processes in diverse porous media. 

Our modeling investigations suggest a nonmonotonic and/or nonlinear dependence of two-phase flow properties on the interfacial tension (or surfactant/PFAS type and concentration) and pore structures (i.e., pore geometry and size distribution). These new two-phase flow properties are validated by a close match between model-predicted and experimentally measured capillary pressure-saturation curves for surfactant-laden fluids in real porous media samples (e.g., quartz sands). Furthermore, these two-phase flow properties are coupled into a general two-phase flow model for simulating the transport of example surfactant-like contaminants (PFAS) through unsaturated soil columns (e.g., the shallow soil layer of the vadose zone). The model simulations reveal that PFAS downward migration is significantly delayed in unsaturated soils with cylindrical pores by the lower water saturation and higher air--water interfacial area, whereas angular pores enhance PFAS retention by increasing the water saturation and decreasing the air--water interfacial area. The findings highlight the critical role of pore angularity on PFAS retention behaviors. 

Beyond the demonstrated PFAS systems, this framework provides a generalizable tool for bridging pore-scale physics and continuum-scale modeling to support predictions and management of multiphase systems in environmental and energy contexts, with potential applications in contaminant transport, enhanced oil recovery, underground CO$_2$/H$_2$ storage, and other relevant problems.

\section{Acknowledgment}
The authors thank Mr. Wenqian Zhang and Ms. Shujie Guo at the University of Arizona for the discussion on the PFAS-modified interfacial tension and contact angle in a single pore.

\appendix
\setcounter{figure}{0} 

\section{Permeability of fluid in a half corner of an angular pore}

The dimensionless conductance of the wetting phase at the $i^{th}$ half corner $\tilde{g}_{\alpha,i}$ given by
\begin{equation}
    \tilde{g}_{\alpha,i} = \exp\left(\frac{m_1\tilde{G}_{\alpha,i}^{2}+m_2\tilde{G}_{\alpha,i} + m_3 + 0.02\sin(\beta_i -\pi/6)}{1/4/\pi-\tilde{G}_{\alpha,i}} +2\ln \tilde{A}_{\alpha,i} \right),
\label{eq:LocalRule_2phase_Krw_g}
\end{equation}
where $m_1 = -18.2066$, $m_2 = 5.88287$, and $m_3 = -0.351809$ are fitting parameters \citep{patzek2001shape},  $\tilde{A}_{\alpha, i}$ and $\tilde{G}_{\alpha, i}$ are the area and shape factor of the wetting-phase region at a corner with a unit meniscus-apex distance (i.e., $l_{\alpha,i}=1$), respectively. $\tilde{A}_{\alpha, i}$ is given by
\begin{equation}
\tilde{A}_{\alpha, i} = \left[\frac{\sin \beta_i}{\cos(\theta+\beta_i)}\right]^2\left[\frac{\cos \theta \cos(\theta+\beta_i)}{\sin \beta_i}+\theta + \beta_i - \frac{\pi}{2}\right],
\label{eq:LocalRule_2phase_Krw_Aw}
\end{equation}
and $\tilde{G}_{\alpha, i}$ is given by
\begin{equation}
\tilde{G}_{\alpha, i} = \frac{\tilde{A}_{\alpha, i} }{4\left[1-(\theta+\beta_i-\pi/2)\sin \beta_i /\cos(\theta_i+\beta)\right]^2}.
\label{eq:LocalRule_2phase_Krw_Gw}
\end{equation}

\section{Pore size distributions}
\begin{figure}[!htb]
  \centering
  \includegraphics[width=0.75\textwidth]{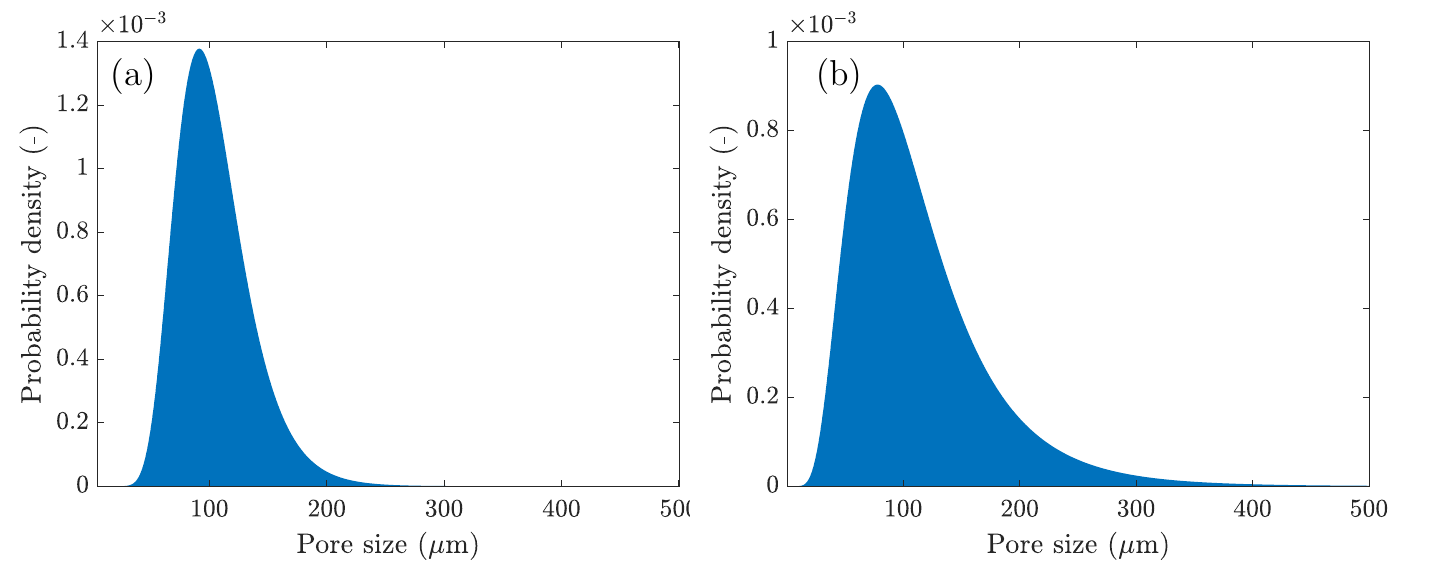}
  \caption{Pore size distributions of porous media with (a) a mean pore size of 100$\mu$m and normalized standard deviation of 0.3, and (b) a mean pore size of 100$\mu$m and normalized standard deviation of 0.5.}
  \label{fig:psd}
\end{figure}

Figure \ref{fig:psd} shows the pore size distributions for the two porous media simulated in Sections \ref{subsec:Impact2Properties} and \ref{subsec:ExplicitEXPvsClosedForm}.

\section{Relative permeability and fluid--fluid interfacial area as functions of saturation in porous media with cylindrical pores}
\begin{figure}[!htb]
  \centering
  \includegraphics[width=0.75\textwidth]{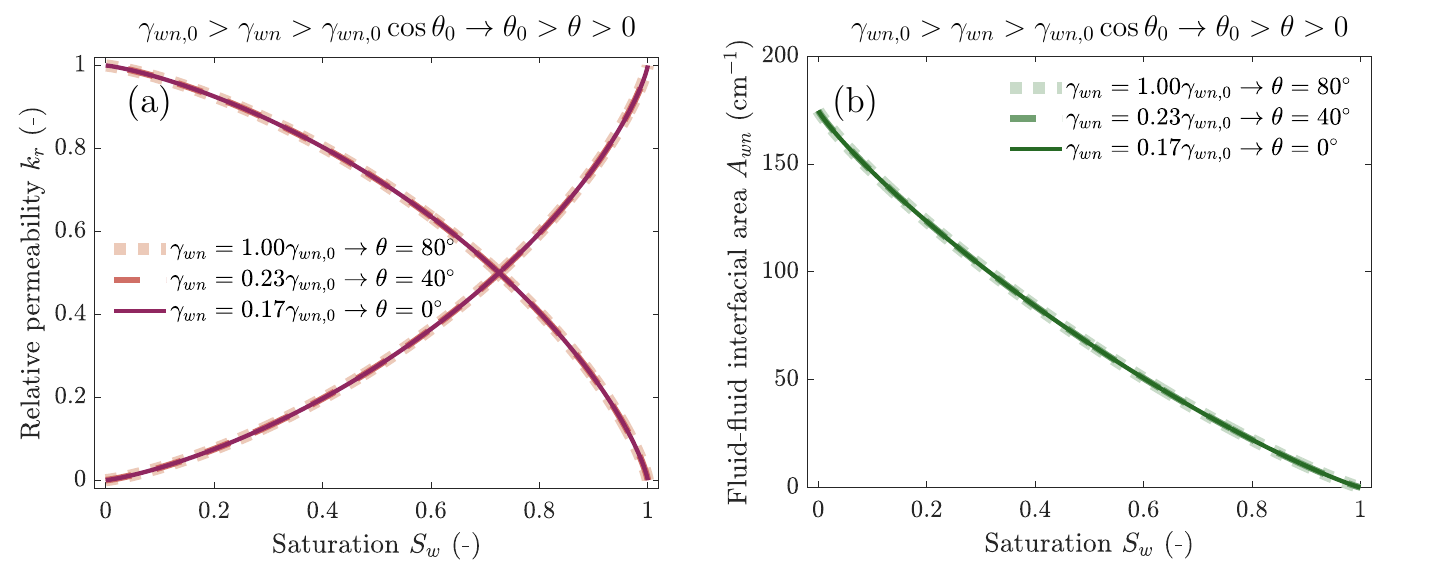}
  \caption{Impact of interfacial tension ($\gamma_{wn}$) on the relative permeability and fluid--fluid interfacial area as functions of saturation in porous media with cylindrical pores. A near neutral-wet condition (e.g., an intrinsic contact angle $\theta_0$ of $80^{\circ}$) is used as an illustrative example.We present the curves for the $\gamma_{nw}$ range where the contact angles become 0 and remain constant. Note that $\gamma_{wn}$ is a proxy of the surfactant effect, i.e., a smaller $\gamma_{wn}$ corresponds to a more interfacially active surfactant and/or a higher surfactant concentration.}
  \label{fig:pc-sw-cylinderical-pores}
\end{figure}

Figure \ref{fig:pc-sw-cylinderical-pores} presents the relative permeability and fluid--fluid interfacial area as functions of saturation in the porous medium with cylindrical pores simulated in \ref{subsec:Impact2Properties}.

\section{Accuracy of the explicit and closed-form expressions}
\begin{figure}[!htb]
  \centering
  \includegraphics[width=1.0\textwidth]{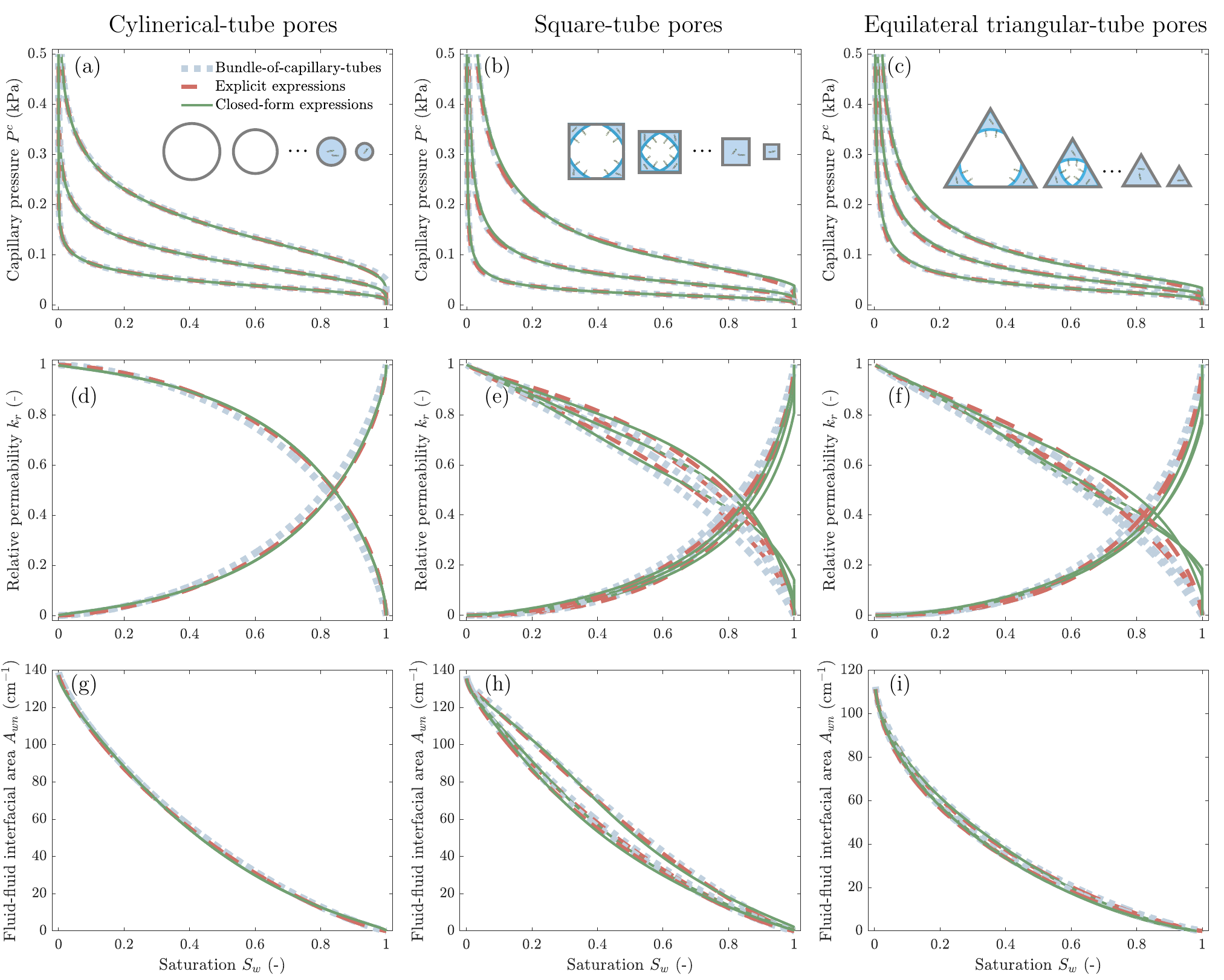}
  \caption{Evaluations on the explicit and closed-form expressions for (a--c) capillary pressure--saturation curves, (e--f) relative permeability--saturation curves, and (g--h) fluid-fluid interfacial area--saturation curves using the bundle-of-capillary-tubes model as a benchmark. Porous media with cylindrical, square-tube, and equilateral triangular-tube pores are examined. A near neutral-wet condition (e.g., an intrinsic contact angle of $80^{\circ}$) is used as an example. We present the curves at three example interfacial tensions ($\gamma_{wn}$) for each pore geometry---i.e., $\gamma_{wn}/\gamma_{wn,0} = 0.22$, $0.1$, and $0.05$ (i.e., $\theta = 40^{\circ}$, $0^{\circ}$ and $0^{\circ}$) for cylindrical pores; $\gamma_{wn}/\gamma_{wn,0} = 0.41$, $0.18$, and $0.05$ (i.e., $\theta = 65^{\circ}$, $20^{\circ}$ and $0^{\circ}$) for square-tube pores; and $\gamma_{wn}/\gamma_{wn,0} = 0.51$, $0.25$, and $0.05$ (i.e., $\theta = 70^{\circ}$, $45^{\circ}$ and $0^{\circ}$) for equilateral triangular-tube pores. A full lognormal pore size distribution is used. The mean and normalized standard deviation of the pore sizes are $100\;{\rm \mu m}$ and $0.5$, respectively.}
  \label{fig:explict-exp-closed-form-0.5-log-normal}
\end{figure}

Figure \ref{fig:explict-exp-closed-form-0.5-log-normal} presents the evaluations of explicit and closed-form expressions for the two-phase flow properties by comparing with the bundle-of-capillary-tubes model.

\section{Temporal and spatial variations of flow and transport variables}
\begin{figure}[!htb]
  \centering
  \includegraphics[width=1.0\textwidth]{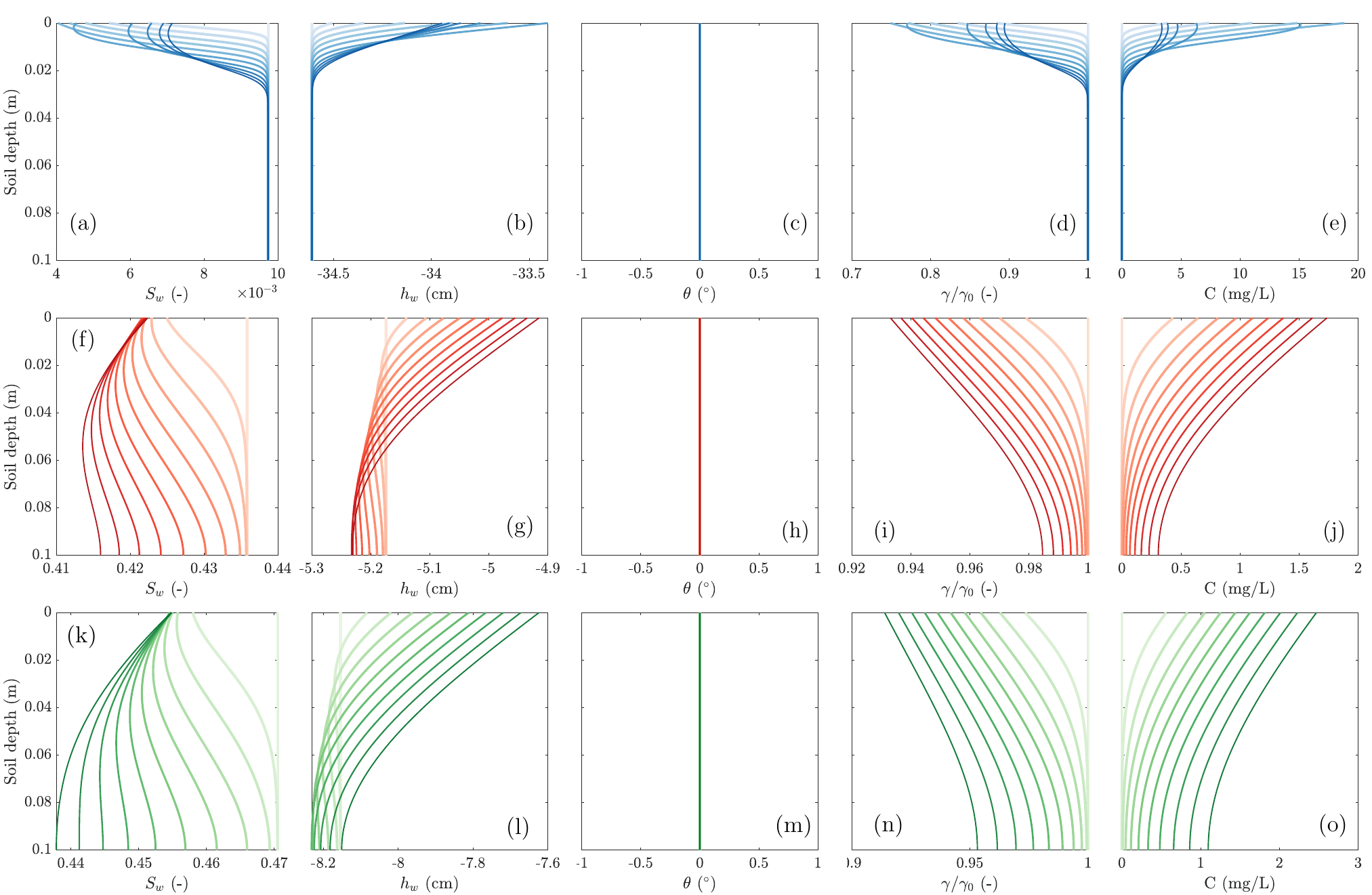}
  \caption{Temporal and spatial variations of water saturation ($S_w$), water pressure head ($h_w$), contact angle ($\theta$), interfacial tension ($\gamma$, which is normalized by the interfacial tension for pure water $\gamma_0$), and aqueous PFOS concentration ($C_{aq}$) in an unsaturated soil column. We present the simulation results for all three pore geometries---including cylindrical tubes (top row), square tubes (middle row), and triangular tubes (bottom row)---at the PFOS concentration of 100\,mg$/$L under completely water-wet condition ($\theta_0 = 0^{\circ}$). We present the profiles from 0 to 4\,hours at an interval of 0.4\,hour for the porous medium with cylindrical pores, and those from 0 to 2\,hours at an interval of 0.2\,hour for the porous medium with angular pores. The lines become progressively darker to represent advancing simulation time.}
  \label{fig:PFAS-profile-comlete-wet}
\end{figure}

\begin{figure}[!htb]
  \centering
  \includegraphics[width=1.0\textwidth]{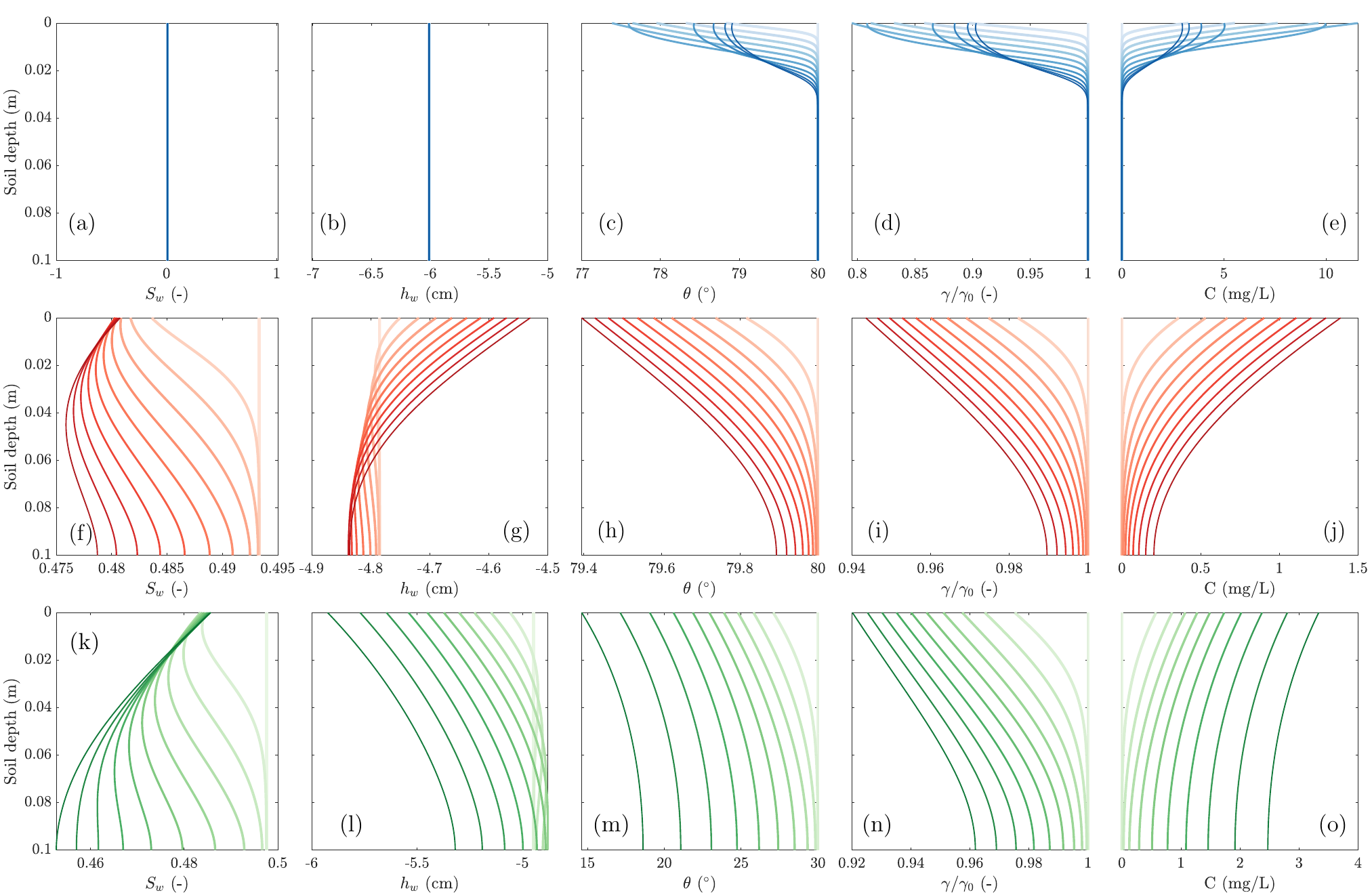}
  \caption{Temporal and spatial variations of water saturation ($S_w$), water pressure head ($h_w$), contact angle ($\theta$), interfacial tension ($\gamma$, which is normalized by the interfacial tension for pure water $\gamma_0$), and aqueous PFOS concentration ($C_{aq}$) in an unsaturated soil column. We present the simulation results for all three pore geometries---including cylindrical tubes (top row), square tubes (middle row), and triangular tubes (bottom row)---at the PFOS concentration of 100\,mg$/$L under weakly water-wet condition ($\theta_0 = 80^{\circ}$) . We present the profiles between 0 and 4\,hours at an interval of 0.4\,hour for the porous medium with cylindrical pores, and those between 0 and 2\,hours at an interval of 0.2\,hour for the porous medium with angular pores. The lines become progressively darker to represent advancing simulation time.}
  \label{fig:PFAS-profile-weak-wet}
\end{figure}

\section{Parameters for experimental validation}

\begin{table}[!htb]
  \centering
  \renewcommand{\arraystretch}{0.7} 
  \setlength{\extrarowheight}{0pt} 
  \setlength{\tabcolsep}{0pt} 
  \caption{Van Genuchten (VG) model parameters for deriving the pore size distribution of F-70 Ottawa sand ($\theta_r$, $\theta_s$, $n$, and $\alpha_{\rm VG}$) and parameters predicting capillary pressure-saturation curves of surfactant-laden fluids. All the parameters were reported in the work of \citet{karagunduz2001influence}. Note that the VG parameters for a surfactant concentration of 0.075\,g/L were mistakenly assigned to 0.050\,g/L in the origianl reference. The corrected values for 0.075\,g/L are provided here.}
  \vspace{5pt}
  \small
  \begin{tabularx}{\textwidth}{lCCC} 
     \hline   
     Surfactant concentration  & Parameter & Unit & Value\\
     \hline
     \multirow{6}{*}{0\,g/L} & $\theta_s$ & cm$^3$/cm$^3$ & 0.3520\\
                             & $\theta_r$ & cm$^3$/cm$^3$ & 0.00008  \\
                             & $n$ & - & 3.6762\\
                             & $\alpha_{\rm VG}$ & Pa$^{-1}$ &  4.1966 \\
                             & $\gamma$ & N/m & 0.072 \\
                             & $\theta$ & $^{\circ}$  & 40.7 \\
    \hline
     \multirow{6}{*}{0.075\,g/L} & $\theta_s$ & cm$^3$/cm$^3$ & 0.3570\\
                                 & $\theta_r$ & cm$^3$/cm$^3$ & 0.00008  \\
                                 & $n$ & - & 4.5561  \\
                                 & $\alpha_{\rm VG}$ & Pa$^{-1}$ & 4.7688  \\
                                 & $\gamma$ & N/m & 0.0379\\
                                 & $\theta$ & $^{\circ}$  & 16.2 \\
    \hline
     \multirow{6}{*}{0.15\,g/L}  & $\theta_s$ & cm$^3$/cm$^3$ & 0.3495\\
                                 & $\theta_r$ & cm$^3$/cm$^3$ & 0.00007 \\
                                & $n$ & - & 3.6070  \\
                                & $\alpha_{\rm VG}$ & Pa$^{-1}$ & 6.0542 \\
                                & $\gamma$ & N/m & 0.0335\\
                                & $\theta$ & $^{\circ}$  & 13.1\\
    \hline   
  \end{tabularx}
  \label{tab:expt_measurement}
\end{table}

\bibliographystyle{elsarticle-harv} 
\bibliography{references}

\end{document}